\documentclass[11pt,pre]{revtex4}
\usepackage[latin9]{inputenc}
\usepackage{amsmath}
\usepackage{graphicx}
\usepackage[FIGTOPCAP]{subfigure}
\usepackage{caption}
\usepackage{amssymb}
\usepackage{esint}
\usepackage{ulem}
\usepackage{cancel}
\usepackage{amsfonts}
\usepackage[unicode=true,pdfusetitle,
 bookmarks=true,bookmarksnumbered=false,bookmarksopen=false,
 breaklinks=false,pdfborder={0 0 1},backref=false,colorlinks=false]
 {hyperref}
\usepackage{color}
\usepackage{setspace}

\makeatletter
\@ifundefined{textcolor}{}
{%
 \definecolor{BLACK}{gray}{0}
 \definecolor{WHITE}{gray}{1}
 \definecolor{RED}{rgb}{1,0,0}
 \definecolor{GREEN}{rgb}{0,1,0}
 \definecolor{BLUE}{rgb}{0,0,1}
 \definecolor{CYAN}{cmyk}{1,0,0,0}
 \definecolor{MAGENTA}{cmyk}{0,1,0,0}
 \definecolor{YELLOW}{cmyk}{0,0,1,0}
}

\usepackage{bm}\usepackage{epsfig}\usepackage{color}

\makeatother

\begin{document}
\singlespacing

\title{Path Integral Formulation for Lévy Flights - Evaluation of the Propagator
for Free, Linear and Harmonic Potentials in the Over- and Underdamped
Limits }
\begin{abstract}
 Lévy flights can be described using a
Fokker-Planck equation, which involves a fractional derivative operator in the position co-ordinate. Such an operator has its natural expression in the Fourier domain. Starting with this, we show that the solution of the equation can be written as a Hamiltonian path integral.   Though this has been realized in the literature, the method has not found applications as the path integral appears difficult to evaluate. We show that a method in which one integrates over the position co-ordinates first, after which integration is performed over the momentum co-ordinates, can be used to evaluate several path integrals that are of interest.  Using this, we evaluate the propagators for  (a) free particle (b) particle subjected to a linear potential
and (c) harmonic potential. In all the three cases, we have obtained results for both overdamped and underdamped cases.

\end{abstract}

\author{Deepika Janakiraman and K.L. Sebastian\\
 Department of Inorganic and Physical Chemistry\\
 Indian Institute of Science\\
 Bangalore 560012\\
 India}

\maketitle

\section{Introduction}

\label{Introduction}

The field of anomalous transport has gained widespread interest due
to its potential to explain several phenomena which fall outside the
realm of simple Brownian motion \cite{ScherLax,Gu,Muller,Klemm,Amblard,GMVish,Reynolds,Atkinson,Ramos,Bartumeus1,Bartumeus2,Sokolov,Brockmann,Lomholtpolymer,Blumen,Levy_glass,Zumofen2}.
The continuous time random walk (CTRW) is a model used to understand anomalous
transport \cite{MontrollWeiss,MontrollScher,physrepMet,Schlesinger1}
where jump lengths and waiting time distributions are both drawn from
a probability distribution function $\psi(x,t)$. The jump length
distribution is given by
\begin{equation}
\lambda(x)=\int_{0}^{\infty}dt\mbox{ }\psi(x,t),\label{jump_length}
\end{equation}
and the waiting time distribution is given by
\begin{equation}
w(t)=\int_{-\infty}^{\infty}dx\mbox{ }\psi(x,t).\label{jump_length}
\end{equation}
Lévy flights belong to a special class of CTRW \cite{PLevy,LevySchlesinger,Chechkin,Encyclopedia,physrepMet},
for which, $\psi(x,t)=\lambda(x)w(t)$. The waiting time distribution, $w(t)$, is narrow, and this makes the process Markovian in nature \cite{Chechkin}. The probability distribution for the jump length, $\lambda(x)$, is  L\'{e}vy stable.  The easiest way to  specify it is to use its  Fourier transform. The most general representation of the Fourier transform of a L\'{e}vy stable distribution is given by \cite{Chechkin},
\begin{equation}
\tilde{L}_{\alpha,\beta}(p;\mu,\sigma)=e^{-\sigma^{\alpha} |p|^{\alpha}\left(1-i\beta \frac{p}{|p|}\omega(p,\alpha)\right)+i\mu p},\label{charfunction_levy}
\end{equation}
where,
\begin{eqnarray}
\omega(p,\alpha)&=&\tan\left(\frac{\pi\alpha}{2}\right), \mbox{  if, }\alpha\neq1\nonumber\\
&=&-\frac{2}{\pi}\ln|p|, \mbox{  if, }\alpha=1
\end{eqnarray}
and $0<\alpha<2$ is the L\'{e}vy index, $-1\leq\beta\leq1$ is the skewness parameter, $\mu\in\mathbb{R}$ is the shift parameter, and $\sigma\in\mathbb{R}^+$ determines the strength of the noise. We will write the position representation for a L\'{e}vy stable distribution as $L_{\alpha,\beta}(x;\mu,\sigma)$. This distribution has the property that it can be rewritten as \cite{Chechkin}
\begin{equation}
\label{levy_rewritten}
L_{\alpha,\beta}(x;\mu,\sigma)=\frac{1}{\sigma}L_{\alpha,\beta}\left(\frac{x-\mu}{\sigma};0,1\right).
\end{equation}
The shift and strength parameters ($\mu,\sigma$) have been absorbed into just one term in the argument. We will adopt an easy to write notation for the L\'{e}vy stable distribution, which is, $\frac{1}{\sigma}L_{\alpha,\beta}\left(\frac{x-\mu}{\sigma}\right)$. If we consider the such a jump length distribution, then the Fourier transform of $\lambda (x)$ is
\begin{equation}
\tilde{\lambda}(p)=e^{-\sigma^{\alpha}|p|^{\alpha}}.
\end{equation}
The L\'{e}vy stable distribution has the following long-tailed behavior for large $x$
\begin{equation}
\lambda(x)\sim\sigma|x|^{-1-\alpha},\mbox{ }x\rightarrow\pm\infty.
\label{asymp_behaviour}
\end{equation}
The Lévy stable distribution reduces to a Gaussian for $\alpha=2$.
Unlike the Gaussian distribution, the Lévy stable distribution does not
obey the usual  central limit theorem. However, it obeys the Lévy-Gnedenko
generalized central limit theorem and is stable under addition \cite{Gnedenko,Feller}.

An inverse power-law asymptotic behavior for the jump length distribution [Eq. (\ref{asymp_behaviour})],
leads to a diverging mean square displacement (MSD) at all times:
\begin{equation}
\langle x^{2}(t)\rangle=\infty.\label{msd}
\end{equation}
This is in contrast to Brownian motion where the MSD is finite and
is proportional to time. This is because, the Brownian walker takes
small steps at each interval of time. On the other hand, a Lévy \textit{flier}
takes small steps interrupted by very long jumps (flights) in between.
As a result, the variance in the step size is infinitely large \cite{physrepMet}.
Lévy flights are observed in a variety of phenomena. Search strategies
of bacteria and various birds and animals, where an occasional long
jump interspersed by short steps offers them significant advantages
\cite{GMVish,Reynolds,Atkinson,Ramos,Bartumeus1,Bartumeus2,ForagingGM},
hopping on a polymer chain \cite{Sokolov,Brockmann,Lomholtpolymer},
diffusion in micelles \cite{Blumen}, optical transport in a Lévy
glass \cite{Levy_glass}, and energy diffusion in single-molecule spectroscopy
\cite{Zumofen2} are a few examples. Levy flights have also been used to
model encounters between different species (predator-prey)\cite{goncharenkovicious2010}.

Lévy flights can be described by a generalization of the Fokker-Planck
equation, referred to as the fractional Fokker-Planck equation(FFPE) \cite{MBK_EuroPL,MBK_PRE,MBK_PRL,physrepMet,dubkovthe2009,CalvoFractionalLevyMotion,Ebeling}.
The FFPE for Lévy flight in the presence of an external potential
$V(x)$ is given by
\begin{equation}
\frac{\partial P(x,t)}{\partial t}=\left\{-D\left(-\frac{\partial^{2}}{\partial x^{2}}\right)^{\frac{\alpha}{2}}+\frac{\partial}{\partial x}\frac{V'(x)}{m\gamma}\right\} P(x,t).\label{Pot FFPE}
\end{equation}
In the above, $D$ is the generalized diffusion constant with dimensions
$[D]=\mbox{meter}^{\alpha}\mbox{ }\mbox{sec}^{-1}$, $V(x)$ is the
potential, $\gamma$ is the friction constant, and $m$ is the mass
of the particle. The solutions of the FFPE for free Lévy flight and for
Lévy flight in the presence of a linear potential and a harmonic potential
in the overdamped limit have been given by Jespersen \textit{et al}. \cite{Fogedby}.
In their paper, they use the FFPE in the Fourier domain. For a free
Lévy flight, it is
\begin{equation}
\frac{\partial\tilde{P}_{free}(k,t)}{\partial t}=-D|k|^{\alpha}\tilde{P}_{free}(k,t),\label{Pot FFPE_free1}
\end{equation}
which can be solved to get
\begin{equation}
\tilde{P}_{free}(k,t)=e^{-D|k|^{\alpha}t}\tilde{P}_{free}(k,0).\label{Pot FFPE_free2}
\end{equation}
Imposing the initial condition $P_{free}(x,0)=\delta(x)$ leads to
$\tilde{P}_{free}(k,t)=e^{-D|k|^{\alpha}t}$ which on Fourier transformation
leads to the solution
\begin{equation}
\label{Fogedby_Free_1}
P_{free}(x,t)=\frac{1}{2\pi}\int_{-\infty}^{\infty}dk\mbox{ }e^{-D|k|^{\alpha}t} e^{i k x}=\frac{1}{(D t)^{1/\alpha}}L_{\alpha,0}\left(\frac{x}{(D t)^{1/\alpha}}\right).
\end{equation}

For a Lévy flight subjected to a linear potential of the form $V(x)=-F_{0}x$,
the FFPE is
\begin{equation}
\frac{\partial\tilde{P}_{lin}(k,t)}{\partial t}=\left(-D|k|^{\alpha}-ik\frac{F_{0}}{\gamma m}\right)\tilde{P}_{lin}(k,t)\label{Pot FFPE_linear1}
\end{equation}
and the solution subject to the same initial condition as before is
\begin{equation}
\tilde{P}_{lin}(k,t)=e^{-t\left(ik\frac{F_{0}}{\gamma m}+D|k|^{\alpha}\right)}.\label{Pot FFPE_linear2}
\end{equation}
This leads to
\begin{equation}
\label{Fogedby_linear}
P_{lin}(x,t)=P_{free}\left(x-\frac{F_{0}t}{\gamma m},t\right)=\frac{1}{(D t)^{1/\alpha}}L_{\alpha,0}\left(\frac{x-\frac{F_{0}t}{\gamma m}}{(D t)^{1/\alpha}}\right).
\end{equation}

Finally, for Lévy flight under a harmonic potential of the form $V(x)=\frac{1}{2}\lambda x^{2}$,
the FFPE is
\begin{equation}
\frac{\partial\tilde{P}_{har}(k,t)}{\partial t}=-\frac{\lambda}{\gamma m}k\frac{\partial}{\partial k}\tilde{P}(k,t)-D|k|^{\alpha}\tilde{P}_{har}(k,t).\label{Pot FFPE_harmonic1}
\end{equation}
One now has to employ the method of characteristics \cite{Fogedby}
to obtain the solution
\begin{equation}
\tilde{P}_{har}(k,t)=\exp\left(-\frac{Dm\gamma|k|^{\alpha}}{\alpha\lambda}\left(1-e^{-\alpha\lambda t/\gamma m}\right)\right).\label{Pot FFPE_harmonic2}
\end{equation}
As a result,
\begin{equation}
\label{Fogedby_harmonic}
P_{har}(x,t)=\left(\frac{\alpha\lambda}{D m \gamma\left(1-e^{-\alpha\lambda t/\gamma m}\right)}\right)^{1/\alpha}L_{\alpha,0}\left(\frac{x}{\left(D m \gamma \left(1-e^{-\alpha\lambda t/\gamma m}\right)/(\alpha\lambda)\right)^{1/\alpha}}\right).
\end{equation}

The path integral approach has found extensive applications in the theory of Brownian motion.  As Brownian motion is  a special case of Levy flights considered here, one would expect path integrals to be very useful for Levy flights too. For example, a path integral
approach for CTRW was discussed by Fredriech and Eule \cite{Fredriech}
and by Calvo \textit{et al.} \cite{CalvoFractionalLevyMotion} for fractional
Brownian motion. However, though the path integral for a Levy processes has been given earlier, it has never been used for evaluating a probability distribution associated with these processes.
An interesting extension of quantum mechanics was suggested by Laskin
\cite{laskinfractional2000,LaskinFractionalQM,LaskinPRE,LaskinPRA,LaskinChaos,Laskinnonlinear}. It consists of modifying
the Schroedinger equation so that one has the operator $\left(-\frac{\partial^{2}}{\partial x^{2}}\right)^{\frac{\alpha}{2}}$in
place of the usual operator $-\frac{\partial^{2}}{\partial x^{2}}$.
He has used a path integral approach and has evaluated some of the
associated path integrals. Our approach in the following is similar
to that of Laskin, and can be used to evaluate Laskin-type path integrals. Essentially, we have the same kind of path integral, but our functionals are all real, as we are concerned with real stochastic processes and not a generalization of quantum mechanics.

There has also been an enormous amount of interest in barrier crossing
by a Brownian particle since the seminal work of Kramers (see the
review by Hanggi \textit{et al.} \cite{Kramers50years}). More recently, the
escape of particles acted upon by fractional noise was investigated
in Refs. \cite{Kramers50years,chaudhurythe2008}. Barrier crossing problem
for Levy flights has been discussed by a few authors \cite{dubkovthe2009,chechkinbarrier2007,dybiecescape2007}.

In this paper we will use a Hamiltonian path integral
approach for Lévy flights. This type of path integrals were introduced
in the context of quantum mechanics, long ago by Garrod \cite{Garrod}.
In these, the integration is over all paths in phase space, which
are not continuous. If the action is quadratic in the momentum variable
$p$, then the momentum path integral can be done easily to get an action
containing only the position and its derivative, and one gets the
more common integral over paths in position space. In the case of
path integrals for Levy flights that we consider, the action expressed
in phase space variables is not quadratic in $p$, but it has the
term $\left|p\right|^{\alpha}$, where $\alpha$ is a number
($0<\alpha<2$) [See Eq. (\ref{prob_x})]. Integrals over $p$ then lead to
symmetric L\'{e}vy stable distribution, and hence one has a path integral involving a product
of $N(\rightarrow\infty)$ such distributions, which appears quite complex.  It may be because of this that
the path integrals for these processes has not been  pursued in the literature. We show that an alternate approach
in which, one first integrates over all the intermediate position variables
and then integrates over the momentum variables, is quite powerful and
allows us to evaluate several path integrals.

Lévy flights in the overdamped limit have been explored by other authors
\cite{physrepMet,Fogedby,Srokowski}. Many of the results that we
find in this limit, using path integrals, have been obtained by other
methods. Our method, however, allows us to evaluate the propagator for L\'{e}vy flight in a harmonic potential with a time-dependent force constant which has not been obtained by solving the FFPE. The underdamped limit for a free Lévy flight has also been
studied, but only to a limited extent by earlier authors. Some of the propagators in
the underdamped limit for the special case of Lévy noise with $\alpha=1$
(Cauchy noise) has been obtained by West and Seshadri \cite{Seshadri}
and Garbaczewski and Olkiewicz \cite{Poland}. Srokowski \cite{Srokowski}
considers a case where the process is driven by an Ornstein-Uhlenbeck
equivalent for the Lévy noise and not by the Lévy noise itself. Both additive
and multiplicative noise are considered in his paper. However, in
the underdamped limit, the propagator for only the free Lévy flight
is obtained, that too under certain conditions. Using our strategy,
we are able to obtain the most general propagator in the underdamped
limit for free Lévy flight and for Lévy flight in a linear and a
harmonic potential. Our results reproduce the existing results under
appropriate conditions. Further, the fact that one can evaluate path
integral for a time-dependent harmonic potential exactly, has
prompted us to study ``semiclassical''-like approximations for path
integrals, which can be very useful in the analysis of Kramers-like
barrier crossing problem for Levy flights. We shall explore this
in a forth coming publication \cite{Deepikatobepublished}.

\section{The Hamiltonian Path Integral}

\label{Hamiltonian_PI}

In this section we shall write a path integral expression for
the propagator for Levy flight, subject to an arbitrary potential
$V(x)$. We start with the characteristic functional for a noise $\eta(t)$
defined by
\begin{equation}
\mathcal{P}[p]=\langle e^{i\int_{0}^{T}dtp(t)\eta(t)}\rangle.\label{characteristicfunctional}
\end{equation}
The expectation value in Eq. (\ref{characteristicfunctional}) may
be written as a path integral over the noise variable, $\eta(t)$,
as
\begin{equation}
\mathcal{P}[p]=\int D\eta\mbox{ }e^{i\int_{0}^{T}dtp(t)\eta(t)}P[\eta].\label{characteristicfunctional_PI}
\end{equation}
$P[\eta]$ is the probability density functional for $\eta(t)$. When the L\'{e}vy noise is delta correlated,
the most general form of the characteristic functional for L\'{e}vy flights is \cite{chechkinintroduction2008}
\begin{equation}
\mathcal{P}[p]=e^{\int_{0}^{T}dt\mbox{  }\left\{i\mu p(t)-D|p(t)|^{\alpha}\left(1-i\beta \frac{p(t)}{|p(t)|}\omega(p(t),\alpha)\right)\right\}}\label{Pbar[p]}
\end{equation}
where, $\alpha$, $\beta$, $\mu$, and $\omega(p(t),\alpha)$ are as defined for Eq. (\ref{charfunction_levy}). $D$ plays the role of determining the noise intensity. Our notation needs a word of explanation. If the functional dependence is shown within usual brackets as in
$P_{free}(x,t)$ in Eq. (\ref{Fogedby_Free_1}), then it is just a function of the variables within the bracket. On the other hand, if we show the independent variable  within  square brackets as in $\mathcal{P}[p]$ of Eq. (\ref{Pbar[p]}), then this means that $\mathcal{P}$ is a functional of the function $p(t)$.  Also, note that $P$ is used to denote the probability density functional, while $\mathcal{P}$ is used for its characteristic functional.   We will describe our procedure for the simplest form of the characteristic functional with $\mu=\beta=0$. However, our method will be applicable even for the general form of the characteristic functional. We will take $D$ to have the dimensions $[D]=\mbox{meter}^{\alpha}/\mbox{sec}$, with $[x]=\mbox{meter}$
and that of $[p]=\mbox{meter}^{-1}$. $P[\eta]$ may be expressed in terms $\mathcal{P}[p]$ as
\begin{equation}
P[\eta]=\int Dp\mbox{ }e^{-i\int_{0}^{T}dt\mbox{ }p(t)\eta(t)}e^{-D\int_{0}^{T}dt|p(t)|^{\alpha}}.\label{prob_noise-1}
\end{equation}
We shall make use of the discretized version of the above integral,
which may be written as
\begin{equation}
P(\eta_{1},\eta_{2}...\eta_{N})=\left(\frac{1}{2\pi}\right)^{N}\int_{-\infty}^{\infty}dp_{1}\int_{-\infty}^{\infty}dp_{2}.....\int_{-\infty}^{\infty}dp_{N}\exp\left(-\Delta t\sum_{n=1}^{N}\left\{ ip_{n}\eta_{n}+D\left|p_{n}\right|^{\alpha}\right\} \right)
\end{equation}
where we have divided the time interval $(0,T)$ into $N(\rightarrow\infty)$
equal intervals, each of length $\Delta t$ and also, $t_{n}=n\Delta t$ and
$\eta_{n}=\int_{t_{n-1}}^{t_{n}}dt\mbox{  }\eta(t)$. $\eta_{n}$ are independent
Levy variables with the probability distribution for $\eta_{n}$ being
given by $P(\eta_{n})=\frac{1}{2\pi}\int_{-\infty}^{\infty}dp_{n}\mbox{  }\exp(-D\Delta t\left|p_{n}\right|^{\alpha}-ip_{n}\eta_{n}\Delta t)$.

The quantity that we are interested in is the probability density
that a particle of mass $m$ executing a Levy flight and starting
at $x_{0}$ at the time $t=0$ would reach $x_{f}$ at the time $t=T$.
The particle is assumed to obey the stochastic differential equation
\begin{equation}
\frac{dx}{dt}+\frac{V'(x)}{m\gamma}=\eta(t).\label{noise_overdamped_dimension}
\end{equation}
$\eta(t)$ is assumed to be a white Levy process, having the characteristic functional $\mathcal{P}[p]=e^{-D\int_0^T dt |p(t)|^\alpha}$, which specifies all its properties.   Using the above equation, and Eq. (\ref{prob_noise-1}), we can write the probability density functional for $x(t)$ as
\begin{equation}
P[x]=\int Dp\: J\mbox{ }e^{-D\int_{0}^{T}dt\mbox{ }|p(t)|^{\alpha}}e^{-i\int_{0}^{T}dt\mbox{ }p(t)\left(\dot{x}+\frac{V'(x)}{m\gamma}\right)}.\label{prob_x}
\end{equation}
$J$ is the Jacobian involved in the transformation from $\eta(t)$
to $x(t)$. The propagator $P(x_{f},T;x_{0},0)$ can now be written
as
\begin{equation}
P(x_{f},T;x_{0},0)=\int_{x(0)=x_{0}}^{x(T)=x_{f}}Dx\int Dp\: J\mbox{ }e^{-D\int_{0}^{T}dt\mbox{ }|p(t)|^{\alpha}}e^{-i\int_{0}^{T}dt\mbox{ }p(t)\left(\dot{x}+\frac{V'(x)}{m\gamma}\right)}.\label{prob_x_conditional}
\end{equation}
At this point, it should be noted that the above equation resembles
the Hamiltonian path integral of quantum mechanics, which was introduced
long ago by Garrod \cite{Garrod}. It is better if one also introduces
the discretized version of the above path integral, so that it is possible to
check for the correctness of the numerical factors, if necessary.
It is convenient to take the discretized version of Eq. (\ref{noise_overdamped_dimension})
to be given by
\begin{equation}
\frac{x_{n}-x_{n-1}}{\Delta t}+\frac{V'(x_{n-1})}{m\gamma}=\eta_{n}.\label{noise_1}
\end{equation}
With this choice of discretization, the Jacobian for the transformation
$J$ becomes unity, and the discretized version reads
\begin{eqnarray}
P(x_{f},T;x_{0},0) & = & \lim_{N\to\infty}\left(\frac{1}{2\pi}\right)^{N}\int_{-\infty}^{\infty}dx_{1}..\int_{-\infty}^{\infty}dx_{N-1}\mbox{ }\int_{-\infty}^{\infty}dp_{1}..\int_{-\infty}^{\infty}dp_{N}\mbox{ }e^{-D\Delta t\sum_{n=1}^{N}|p_{n}|^{\alpha}}\times\nonumber\\
 &  & e^{-i\sum_{n=1}^{N}p_{n}\left(\frac{x_{n}-x_{n-1}}{\Delta t}+\frac{V'(x_{n-1})}{m\gamma}\right)\Delta t}.
\label{eq:DiscretizedVersion}
\end{eqnarray}

Barkai \textit{et al.} \cite{BarkaiJStatPhys2010} have derived a fractional
generalization of the Feynman-Kac equation for functionals of
sub-diffusive continuous-time random walks. They have also derived a backward equation and a generalization
to L\'{e}vy flights. Solutions are presented for a wide number of applications including
the occupation time in half space and in an interval, the first passage time, the maximal
displacement, and the hitting probability. For the particular class of processes that we are considering, it is quite easy to derive such an equation [Eq. (\ref{Pot FFPE}) of this paper] using the above path integral.  The details of this calculation is given in Appendix A.
In the following, we shall write Eq. (\ref{eq:DiscretizedVersion}) as
\begin{equation}
P(x_{f},T;x_{0},0)=\int_{x(0)=x_{0}}^{x(T)=x_{f}}Dx\mbox{ }\int Dp\mbox{ }e^{-D\int_{0}^{T}dt\mbox{ }|p(t)|^{\alpha}}e^{-i\int_{0}^{T}dt\mbox{ }p(t)\left(\dot{x}+\frac{V'(x)}{m\gamma}\right)}.\label{eq:pathintegral}
\end{equation}

Note that we have adopted the Ito discretization, which is very convenient
for the problem as there is no contribution to the Jacobian from the potential, $V(x)$, in Eq. (\ref{noise_1}). The alternate Stratanovich version of discretization is
\begin{equation}
\frac{x_{n}-x_{n-1}}{\Delta t}+\frac{V'(x_{n})+V'(x_{n-1})}{2m\gamma}=\eta_{n}\label{noise_discretized_Stratnovich}
\end{equation}
and results in a more complicated expression for the Jacobian
\cite{Bray}.

If $\alpha=2,$ then the integrals over $p_{n}$s in Eq. (\ref{eq:DiscretizedVersion})
are simple Gaussian integrals and can be done easily to get the usual
path integrals for Brownian motion. For other values of
$\alpha$, each integral over $p_{n}$ will result in a stable distribution  of the form
\begin{equation}
\int_{-\infty}^{\infty}dp_{n}\mbox{ }e^{-D\Delta t|p_{n}|^{\alpha}} e^{-ip_{n}\left(x_{n}-x_{n-1}+\frac{V'(x_{n-1})}{m\gamma}\Delta t\right)}=\frac{2\pi}{(D \Delta t)^{1/\alpha}} L_{\alpha,0}\left(\frac{x_{n}-x_{n-1}+\frac{V'(x_{n-1})}{m\gamma}\Delta t}{(D \Delta t)^{1/\alpha}}\right).\end{equation}

Therefore, Eq. (\ref{eq:DiscretizedVersion}) will become a product
of $N$ such stable distributions, where $N\to\infty$ and this product
has to be integrated over $x_{n}$, which seems formidable. This
is probably the reason why a path integral approach for Lévy flights
has been used rarely in the literature.

Our approach to the path integral in Eq. (\ref{eq:pathintegral}) is simple. If $V(x)$ is at the
most a quadratic function of $x,$ then it is possible to first integrate
over the position co-ordinates exactly leaving us with a product of
Dirac delta functions involving the $p_{n}$s. The integrals over $p_{n}$
can then be performed easily to get the final answer. Using the procedure
delineated here, we will obtain $P(x_{f},T;x_{0},0)$ for the free
Lévy flight and Lévy flight in the presence of linear and harmonic
potentials and thereby, reproduce the results of Jespersen \textit{et al}.
\cite{Fogedby}. For the harmonic potential, we will present results
for both time-independent as well as time-dependent force constants.
The result for time-independent force constant is given in the paper
by Jespersen \textit{et al}. \cite{Fogedby}. However, the propagator for the Lévy
flight in a harmonic potential with a time-dependent force constant
is a new result. Also, using the same approach, we will obtain the most general propagator in the
underdamped limit for free Lévy flight and for Lévy flight in a linear potential and a harmonic
potential with a time-independent force constant, which are also new
results. The procedure can also be applied to find ``semiclassical''
approximations for more complicated $V(x)$ and can be used to analyze
the Kramers problem for a Levy particle \cite{Deepikatobepublished}.  We adopt the following notations: In the overdamped limit, the propagator will have the superscript
`od' ($P^{od}$), and in the underdapmed limit, it has the superscript
`ud' ($P^{ud}$). We also write $p(T)=p_T$ and $p(0)=p_0$.

\section{Application to Various Potentials in the Overdamped limit}

\label{overdamped_section}

\subsection{Free Lévy flight}

\label{Free_Levy_over}
\noindent
For a free particle, $V(x)=0$. For this case, Eq. (\ref{eq:DiscretizedVersion})
becomes
\begin{eqnarray}
P_{free}^{od}(x_{f},T;x_{0},0) & = & \lim_{N\to\infty}\left(\frac{1}{2\pi}\right)^{N}\int_{-\infty}^{\infty}dp_{1}..\int_{-\infty}^{\infty}dp_{N}\mbox{ }e^{-D\Delta t\sum_{n=1}^{N}|p_{n}|^{\alpha}}e^{-ip_{N}x_{f}}e^{ip_{1}x_{0}}\times\nonumber \\
 &  & \prod_{n=1}^{N-1}\int_{-\infty}^{\infty}dx_{n}\mbox{ }e^{i(p_{n+1}-p_{n})x_{n}}.
\end{eqnarray}
After performing integrals over all $x_{i}$s, the resulting expression
is
\begin{eqnarray}
P_{free}^{od}(x_{f},T;x_{0},0) & = & \lim_{N\to\infty}\frac{1}{2\pi}\int_{-\infty}^{\infty}dp_{N}\mbox{ }e^{-D|p_{N}|^{\alpha}\Delta t}e^{-ip_{N}x_{N}}\left(\prod_{n=1}^{N-1}\int_{-\infty}^{\infty}dp_{n}\mbox{ }e^{-D|p_{n}|^{\alpha}\Delta t}\delta(p_{n+1}-p_{n})\right)e^{ip_{1}x_{0}}. \nonumber\\
\end{eqnarray}
Integrating over $p_{n}$ with $n=1$ to $N-1$ gives
\begin{equation}
P_{free}^{od}(x_{f},T;x_{0},0)=\frac{1}{2\pi}\int_{-\infty}^{\infty}dp_{N}\mbox{ }e^{-DT|p_{N}|^{\alpha}}e^{-ip_{N}(x_{f}-x_{0})},\label{prob_x_free3}
\end{equation}
which leads to \cite{Fogedby}
\begin{equation}
\label{prob_x_fox-H}
P_{free}^{od}(x_{f},T;x_{0},0)=\frac{1}{{(DT)^{1/\alpha}}}L_{\alpha,0}\left(\frac{x_{f}-x_{0}}{(DT)^{1/\alpha}}\right).
\end{equation}

In the following, one can proceed with similar discretized integrals
to evaluate the propagator for other problems. The discretized version
of the path integral is tedious to perform, and it is far easier if
one adopted a continuum version of the same integral. To illustrate
the approach, we will do the above integral using the continuum approach.
The approach has the difficulty that the propagator is determined
only to within a multiplicative factor. However, this is not a problem
as one can determine the factor using the normalization condition
on the propagator. In all the further calculations, we will use the
continuum version only. The propagator for free Lévy flight in the
continuum version is
\begin{equation}
P_{free}^{od}(x_{f},T;x_{0},0)=\int_{x(0)=x_{0}}^{x(T)=x_{f}}Dx\int Dp\mbox{ }e^{-D\int_{0}^{T}dt\mbox{ }|p(t)|^{\alpha}}e^{-i\int_{0}^{T}dt\mbox{ }p(t)\dot{x}(t)}.\label{free_continuum}
\end{equation}
On performing the integral in the exponent, $\int_{0}^{T}dt\mbox{ }p(t)\dot{x}(t)$, by parts, we get
\begin{eqnarray}
P_{free}^{od}(x_{f},T;x_{0},0) & = & \int Dp\mbox{ }e^{-D\int_{0}^{T}dt\mbox{ }|p(t)|^{\alpha}}e^{-i\left(p_{T}x_{f}-p_{0}x_{0}\right)}\int_{x(0)=x_{0}}^{x(T)=x_{f}}Dx\mbox{ }e^{i\int_{0}^{T}dt\mbox{ }\dot{p}(t)x(t)}\nonumber \\
 & = & \int Dp\mbox{ }e^{-D\int_{0}^{T}dt\mbox{ }|p(t)|^{\alpha}}e^{-i\left(p_{T}x_{f}-p_{0}x_{0}\right)}\delta[\dot{p}(t)].
\end{eqnarray}
In the above equation, $\delta[\dot{p}(t)]$ stands for Dirac delta functional, which results from the path integral over $x$. Dirac delta functional implies that
\begin{equation}
\dot{p}(t)=0\mbox{ }\Rightarrow\mbox{ }p(t)=p_{0},\mbox{ a constant}.\label{delta_func_result_free}
\end{equation}
On performing the integral over $p(t)$, taking into  account the  delta functional and determining
the multiplicative factor from the normalization condition, we obtain
\begin{equation}
P_{free}^{od}(x_{f},T;x_{0},0)=\frac{1}{2\pi}\int_{-\infty}^{\infty}dp_{0}\mbox{ }e^{-D|p_{0}|^{\alpha}T}e^{-ip_{0}(x_{f}-x_{0})}=\frac{1}{(DT)^{1/\alpha}}L_{\alpha,0}\left(\frac{x_f-x_0}{(DT)^{1/\alpha}}\right).\label{free_continuum2}
\end{equation}
This is the same result that we obtained using the discretized version
of the path integral in Eq. (\ref{prob_x_fox-H}).

\subsection{Linear potential}
\label{Linear_Levy_over}
We now analyze Lévy flight under the influence of a linear potential
of the form $V(x)=-F_{0}x$. The propagator is
\begin{equation}
P_{lin}^{od}(x_{f},T;x_{0},0)=\int_{x(0)=x_{0}}^{x(T)=x_{f}}Dx\mbox{ }\int Dp\mbox{ }e^{-D\int_{0}^{T}dt\mbox{ }|p(t)|^{\alpha}}e^{-i\int_{0}^{T}dt\mbox{ }p(t)\left(\dot{x}(t)-\frac{F_{0}}{m\gamma}\right)}.\label{linear_continuum}
\end{equation}
On performing the integral in the exponent, $\int_{0}^{T}{dt\mbox{ }p(t)\left(\dot{x}(t)-\frac{F_{0}}{m\gamma}\right)}$,
by parts,
\begin{eqnarray}
P_{lin}^{od}(x_{f},T;x_{0},0) & = & \int Dp\mbox{ }e^{-D\int_{0}^{T}dt\mbox{ }|p(t)|^{\alpha}}e^{-i\left(p_{T}x_{f}-p_{0}x_{0}\right)}e^{i\frac{F_{0}}{m\gamma}\int_{0}^{T}dt\mbox{ }p(t)}\int_{x(0)=x_{0}}^{x(T)=x_{f}}Dx\mbox{ }e^{i\int_{0}^{T}dt\mbox{ }\dot{p}(t)x(t)}\nonumber \\
\end{eqnarray}
Integrating over all the paths in position space with fixed end points,
we get
\begin{equation}
P_{lin}^{od}(x_{f},T;x_{0},0)=\int Dp\mbox{ }e^{-D\int_{0}^{T}dt\mbox{ }|p(t)|^{\alpha}}e^{-i\left(p_{T}x_{f}-p_{0}x_{0}\right)}e^{i\frac{F_{0}}{m\gamma}\int_{0}^{T}dt\mbox{ }p(t)}\delta[\dot{p}(t)].
\end{equation}
The Dirac delta functional $\delta[\dot{p}(t)]$, implies that
\begin{equation}
\dot{p}(t)=0\mbox{ }\Rightarrow\mbox{ }p(t)=p_{0},\mbox{ a constant}.\label{delta_func_result_lin}
\end{equation}
Performing the path integration over $p(t)$ accounting for the delta functional and determining
the multiplicative factor such that $P_{lin}^{od}$ is normalized, gives
\begin{equation}
P_{lin}^{od}(x_{f},T;x_{0},0)=P_{free}^{od}\left(x_{f}-F_{0}T/m\gamma,T;x_{0},0\right).\label{lin_free}
\end{equation}
If $x_{0}=0$, this is just the result of Jespersen \textit{et al}. presented
in Eq. (\ref{Fogedby_linear}).

It may be noted that for $\mu\neq0\mbox{ and }\beta=0$ in Eq. (\ref{Pbar[p]}), the equations will turn out to be exactly the same as L\'{e}vy flight in a linear potential. For the case where $\beta\neq0$, this procedure, where we first integrate over position coordinates first, is still valid. As an example, let us consider free L\'{e}vy flight where $\beta\neq0$ and $\mu\neq0$. After performing the integration over position variables, we will be left with the integral
\begin{equation}
\frac{1}{2\pi}\int_{-\infty}^{\infty}dp_{0}\mbox{ }e^{-D|p_{0}|^{\alpha}T\left(1-\beta\frac{p_0}{|p_0|}\omega(p_0,\alpha)\right)}e^{-ip_{0}(x_{f}-x_{0}-\mu)}=\frac{1}{(DT)^{1/\alpha}}L_{\alpha,\beta}\left(\frac{x_{f}-x_{0}-\mu}{(D T)^{1/\alpha}}\right).
\end{equation}
The final distribution is L\'{e}vy stable, just that it is not symmetric, as expected.

\subsection{Harmonic potential with a time-dependent Force Constant}

\label{Harmonic_Levy_over}

We will now consider Lévy flight in a harmonic potential with a time-dependent force constant, for which the results have not been reported
in the literature. It is interesting to note that our method
is very easy when applied to this problem, while in comparison solving
the corresponding fractional differential equation would be much more
involved. We write the potential as $V(x)=\frac{1}{2}\lambda(t)x^{2}$
and find an analytical expression for the propagator using the Hamiltonian
path integral approach. The path integral for the propagator is
\begin{equation}
P_{har}^{od}(x_{f},T;x_{0},0)=\int_{x(0)=x_{0}}^{x(T)=x_{f}}Dx\mbox{ }\mbox{ }\mbox{ }\int Dp\mbox{ }e^{-D\int_{0}^{T}dt\mbox{ }|p(t)|^{\alpha}}e^{-i\int_{0}^{T}dt\mbox{ }p(t)\left(\dot{x}(t)+\frac{\lambda(t)}{m\gamma}x(t)\right)}.\label{harmonic_continuum}
\end{equation}
On performing the integral in the exponent, $\int_{0}^{T}dt\mbox{ }p(t)\dot{x}(t)$, by parts,
\begin{eqnarray}
P_{har}^{od}(x_{f},T;x_{0}) & = & \int Dp\mbox{ }e^{-D\int_{0}^{T}dt\mbox{ }|p(t)|^{\alpha}}e^{-i\left(p_{T}x_{f}-p_{0}x_{0}\right)}\int_{x(0)=x_{0}}^{x(T)=x_{f}}Dx\mbox{ }\mbox{ }e^{i\int_{0}^{T}dt\mbox{ }\left(\dot{p}(t)-\frac{\lambda(t)}{m\gamma}p(t)\right)x(t)}\nonumber \\
 & = & \int Dp\mbox{ }e^{-D\int_{0}^{T}dt\mbox{ }|p(t)|^{\alpha}}e^{-i(p_{T}x_{f}-p_{0}x_{0})}\delta\left[\dot{p}(t)-\frac{\lambda(t)}{m\gamma}p(t)\right].
\end{eqnarray}
The delta functional leads to the relation
\begin{equation}
\dot{p}(t)-\frac{\lambda(t)}{m \gamma}p(t)=0\mbox{ }\Rightarrow\mbox{ }p(t)=p_{0}e^{\int_{0}^{t}dt'\mbox{ }\lambda(t')/m \gamma}.\label{delta_func_result_har}
\end{equation}
After performing the integration over the delta functional and making a change of integration variable $\tilde{p}_{0}=p_{0}\mbox{ }e^{\int_{0}^{T}dt'\mbox{ }\lambda(t')/m\gamma}$, the propagator, $P_{har}^{od}(x_{f},T;x_{0},0)$ along with the multiplicative factor
is
\begin{equation}
\label{Fox_Harmonic_time_dependent}
P_{har}^{od}(x_{f},T;x_{0})=\frac{1}{\left(D\int_{0}^{T}dt\mbox{ }e^{-\alpha\int_{t}^{T}dt'\mbox{ }\lambda(t')/m \gamma}\right)^{1/\alpha}}L_{\alpha,0}\left(\frac{x_{f}-x_{0}e^{-\int_{0}^{T}dt'\mbox{ }\frac{\lambda(t')}{m \gamma}}}{\left(D\int_{0}^{T}dt\mbox{ }e^{-\alpha\int_{t}^{T}dt'\mbox{ }\lambda(t')/m \gamma}\right)^{1/\alpha}}\right)
\end{equation}
For time-independent force constant, the result in Eq. (\ref{Fox_Harmonic_time_dependent})
reduces to the following
\begin{eqnarray}
P_{har}^{od}(x_{f},T;x_{0})=\frac{1}{\left(D m\gamma(1-e^{-\frac{\alpha\lambda T}{m \gamma}})/(\alpha\lambda)\right)^{1/\alpha}} L_{\alpha,0}\left(\frac{x_{f}-x_{0}e^{-\frac{\lambda T}{m \gamma}}}{\left(D m\gamma(1-e^{-\frac{\alpha\lambda T}{m \gamma}})/(\alpha\lambda)\right)^{1/\alpha}}\right).
\end{eqnarray}
This result is identical to that of Jespersen \textit{et al}. in Eq. (\ref{Fogedby_harmonic})
for the initial condition $x_{0}=0$.  Clearly as $T\rightarrow \infty$ the above function approaches $\frac{1}{\left(Dm\gamma\right)^{1/\alpha}}L_{\alpha,0}\left(\frac{x_f}{\left(Dm\gamma\right)^{1/\alpha}}\right)$, showing that the probability distribution attains a steady value at infinite time. The  average energy of the particle is given by the expression $\frac{1}{2}\lambda \langle x_f^2\rangle$. Since the mean square displacement of a L\'{e}vy distribution always diverges [Eq. (\ref{msd})], the average energy of the particle also diverges.

\section{Lévy Flight in the Underdamped Limit }

The use of our procedure, enables us to obtain the most general propagator in the underdamped limit, which is a new result. In the underdamped limit, the inertial
term in the Langevin equation cannot be ignored and the motion is
governed by the equation
\begin{equation}
\frac{\ddot{x}(t)}{\gamma}+\dot{x}(t)+\frac{V'\left(x(t)\right)}{m\gamma}=\eta(t).\label{underdamped_noise_1}
\end{equation}
Here also, we will use the Hamiltonian path integral formulation in
its continuum version. We will present results for the free Lévy flight
and also in the presence of a linear potential and a harmonic potential with a time-independent
force constant.

\subsection{Free Lévy flight}

\label{Free_Levy_under}

The propagator for free Lévy flight in the underdamped regime is
\begin{equation}
P_{free}^{ud}(x_{f},T;x_{0},0)=\int_{x(0)=x_{0}}^{x(T)=x_{f}}Dx\int Dp\mbox{ }e^{-D\int_{0}^{T}dt\mbox{ }|p(t)|^{\alpha}}e^{-i\int_{0}^{T}dt\mbox{ }p(t)\left(\frac{\ddot{x}(t)}{\gamma}+\dot{x}(t)\right)}.\label{free_underdamp}
\end{equation}
Integrating $\int_{0}^{T}dt\mbox{ }p(t)\left(\frac{\ddot{x}(t)}{\gamma}+\dot{x}(t)\right)$
in the exponent by parts we get
\begin{eqnarray}
\int_{0}^{T}dt\mbox{ }p(t)\left(\frac{\ddot{x}(t)}{\gamma}+\dot{x}(t)\right) & = & \frac{p_{T}v_{f}-p_{0}v_{0}-\dot{p}_{T}x_{f}+\dot{p}_{0}x_{0}}{\gamma}+p_{T}x_{f}-p_{0}x_{0}+\nonumber \\
 &  & \int_{0}^{T}dt\mbox{ }\left(\frac{\ddot{p}(t)}{\gamma}-\dot{p}(t)\right)x(t),
\end{eqnarray}
where, $v_{0}$ and $v_f$ are the initial and final
velocities, respectively. Since the differential equation governing
the motion is second order [Eq. (\ref{underdamped_noise_1})], the
propagator will depend on the initial and final positions as well
as the velocities. The propagator is
\begin{eqnarray}
P_{free}^{ud}(x_{f},v_f,T;x_{0},v_0,0) & = & \int Dp\mbox{ }e^{-D\int_{0}^{T}dt\mbox{ }|p(t)|^{\alpha}}e^{-i\left(\frac{p_{T}v_f-p_{0}v_0-\dot{p}_{T}x_{f}+\dot{p}_{0}x_{0}}{\gamma}+p_{T}x_{f}-p_{0}x_{0}\right)}\times\nonumber \\
 &  & \int Dx\mbox{ }e^{-i\int_{0}^{T}dt\mbox{ }\left(\frac{\ddot{p}(t)}{\gamma}-\dot{p}(t)\right)x(t)}\label{free_underdamp1} \nonumber\\
 & = & \int Dp\mbox{ }e^{-D\int_{0}^{T}dt\mbox{ }|p(t)|^{\alpha}}e^{-i\left(\frac{p_{T}v_f-p_{0}v_0-\dot{p}_{T}x_{f}+\dot{p}_{0}x_{0}}{\gamma}+p_{T}x_{f}-p_{0}x_{0}\right)}\delta\left[\ddot{p}(t)-\gamma\dot{p}(t)\right]. \nonumber
\end{eqnarray}
Dirac delta functional leads to the condition
\begin{equation}
\ddot{p}(t)-\gamma\dot{p}(t)=0,
\end{equation}
which has the solution
\begin{equation}
\therefore p(t)=c_{1}+c_{2}e^{\gamma t},
\end{equation}
where $c_{1}=\frac{p_{T}-p_{0}e^{\gamma T}}{1-e^{\gamma T}}$ and
$c_{2}=\frac{p_{0}-p_{T}}{1-e^{\gamma T}}$. On performing the integral
over the delta functional, the path integral over all $p$s
reduces to just two integrals over $p_{T}$ and $p_{0}$. Therefore, the unnormalized propagator is
\begin{eqnarray}
P_{free}^{ud}(x_{f},v_f,T;x_{0},v_0,0)&=&\int_{-\infty}^{\infty}dp_{0}\int_{-\infty}^{\infty}dp_{T}\mbox{ }e^{-D\int_{0}^{T}dt\mbox{ }\left|p_T\frac{1-e^{\gamma t}}{1-e^{\gamma T}}+p_0\frac{e^{\gamma t}-e^{\gamma T}}{1-e^{\gamma T}}\right|^{\alpha}}\nonumber\\
&&\exp\left[-i\left\{p_T\frac{v_f}{\gamma}-p_0\frac{v_0}{\gamma}+\frac{p_0 e^{\gamma T}-p_T}{e^{\gamma T}-1}(x_f-x_0)\right\}\right].\label{free_underdamp2}
\end{eqnarray}
After normalizing, the propagator becomes
\begin{eqnarray}
P_{free}^{ud}(x_{f},v_f,T;x_{0},v_0,0)&=&\frac{1}{4\pi^2\gamma\left(1-e^{-\gamma T}\right)}\int_{-\infty}^{\infty}dp_{0}\int_{-\infty}^{\infty}dp_{T}\mbox{ }e^{-D\int_{0}^{T}dt\mbox{ }\left|\frac{p_0\left(1-e^{-\gamma t}\right)+p_T\left(e^{-\gamma t}-e^{-\gamma T}\right)}{1-e^{-\gamma T}}\right|^{\alpha}}\times\nonumber\\
&&\exp\left[-i\left\{p_T\frac{v_f}{\gamma}-p_0\frac{v_0}{\gamma}+\frac{p_0-p_T e^{-\gamma T}}{1-e^{-\gamma T}}(x_{f}-x_{0})\right\}\right].\label{free_underdamp2_1}
\end{eqnarray}
For $\alpha=2$, the above expression corresponds
to Brownian motion, the integrals can be performed exactly and, the propagator
we obtain matches with the result given in the seminal paper by Chandrasekhar
\cite{Chandrasekhar}. For $\alpha\neq2$, it seems difficult to evaluate the double integral in Eq. (\ref{free_underdamp2_1}).
In order to simplify the evaluation of the propagator, we make a change
of variables from $\{p_{0},p_{T}\}$ to $\{q_{1},q_{2}\}$, such that
$p_T=q_{1}$ and $p_{0}=q_{1}q_{2}$. The details of this transformation are given in Appendix B. On making the change of variables, we get
\begin{equation}
P_{free}^{ud}(x_{f},v_f,T;x_{0},v_0,0)=\frac{1}{2\pi^2\gamma\alpha\left(1-e^{-\gamma T}\right)}\int_{-\infty}^{\infty}dq_{2}\int_{0}^{\infty}dq_1\mbox{ }q_1e^{-\Theta_1|q_1|^{\alpha}}\cos\left(\Lambda_1 q_1\right)
\label{free_underdamp2_2}
\end{equation}
where,
\begin{eqnarray}
\Theta_1&=&D\int_{0}^{T}dt\mbox{ }\left|\frac{e^{-\gamma t}-e^{-\gamma T}}{1-e^{-\gamma T}}+q_2\frac{1-e^{-\gamma t}}{1-e^{-\gamma T}}\right|^{\alpha},\nonumber\\
\Lambda_1&=&\left(\frac{v_f}{\gamma}-\frac{e^{-\gamma T}(x_{f}-x_{0})}{1-e^{-\gamma T}}\right)-q_2\left(\frac{v_0}{\gamma}-\frac{(x_{f}-x_{0})}{1-e^{-\gamma T}}\right).
\end{eqnarray}
The integral over $q_1$ can be performed exactly for $\alpha=1$ and written as
\begin{equation}
P_{free}^{ud}(x_{f},v_f,T;x_{0},v_0,0)=\frac{1}{2\pi^2\gamma\left(1-e^{-\gamma T}\right)}\int_{-\infty}^{\infty}dq_{2}\mbox{ }\frac{\Theta_1^2-\Lambda_1^2}{\left(\Theta_1^2+\Lambda_1^2\right)^2}.
\label{free_underdamp2_3}
\end{equation}
For other values of $\alpha$, the integral over $q_1$ cannot be performed exactly. However, we expand the cosine function in Eq. (\ref{free_underdamp2_2}) and then perform the integral over $q_1$, which results in
\begin{equation}
\label{most_general_free_sum}
P_{free}^{ud}(x_{f},v_f,T;x_{0},v_0)=\frac{1}{2\pi^2\gamma\alpha\left(1-e^{-\gamma T}\right)}\int_{-\infty}^{\infty}dq_2\sum_{n=0}^{\infty}\frac{(-1)^n}{(2n)!}
\mbox{ }\Gamma\left(\frac{2(n+1)}{\alpha}\right)\frac{\Lambda_1^{2n}}{\Theta_1^{2(n+1)/\alpha}}.
\end{equation}
The sum does not converge for $\alpha<1$. As $T\rightarrow\infty$, $P_{free}^{ud}(x_{f},v_f,T;x_{0},v_0)\rightarrow0$ because $\Theta_1\rightarrow\infty$ in this limit for any non-zero value of $q_2$.

More specific
propagators can be derived from $P_{free}^{ud}(x_{f},v_f,T;x_{0},v_0,0)$.
For the case of Brownian motion, the propagators $P_{free}^{ud}(x_{f},T;x_{0},v_0,0)$
and $P_{free}^{ud}(v_f,T;v_0,0)$ were given in Chandrasekhar's
paper. Similar exact results can be obtained for Lévy flights using
our formalism. The details of the calculation are provided in Appendix
C. The propagator
\begin{eqnarray} P_{free}^{ud}(x_{f},T;x_{0},v_0,0)&=&\int_{-\infty}^{\infty}dv_f\mbox{ }P_{free}^{ud}(x_{f},v_f,T;x_{0},v_0,0)\nonumber\\
 & = & \frac{1}{(1-e^{-\gamma T})\Theta_2^{1/\alpha}}L_{\alpha,0}\left(\frac{\Lambda_2}{\Theta_2^{1/\alpha}}\right),\label{free_underdamp4}
\end{eqnarray}
with
\begin{equation}
\Theta_2=D\int_{0}^{T}dt\mbox{ }\left(\frac{1-e^{-\gamma t}}{1-e^{-\gamma T}}\right)^{\alpha}=\frac{D}{\gamma}\frac{B\left(1-e^{-\gamma T};1+\alpha,0\right)}{\left(1-e^{-\gamma T}\right)^{\alpha}},
\end{equation}
and
\begin{equation}
\Lambda_2=\frac{v_0}{\gamma}-\frac{x_f-x_0}{1-e^{-\gamma T}}.
\end{equation}
In the limit $T\rightarrow\infty$, $\Theta_2\rightarrow\infty$ and therefore, in this limit, $P_{free}^{ud}(x_{f},T;x_{0},v_0,0)\rightarrow0$.

On integrating over all possible values of $x_f$ of $P_{free}^{ud}(v_f,x_{f},T;v_0,0)$,
we get
\begin{eqnarray}
P_{free}^{ud}(v_f,T;v_0,0)&=&\int_{-\infty}^{\infty}dx_f\mbox{ }P_{free}^{ud}(x_{f},v_f,T;x_{0},v_0,0)\nonumber \\
 & = &\frac{1}{\Theta_3^{1/\alpha}}L_{\alpha,0}\left(\frac{\Lambda_3}{\Theta_3^{1/\alpha}}\right),
\end{eqnarray}
where
\begin{equation}
\Theta_3=\frac{D}{\alpha \gamma^{1-\alpha}}\left(1-e^{-\alpha\gamma T}\right),
\end{equation}
and
\begin{equation}
\Lambda_3=v_f-v_0 e^{-\gamma T}.
\end{equation}
The stationary distribution for $P_{free}^{ud}(v_f,T;x_{0},v_0,0)$ is obtained by letting $T \rightarrow \infty$, which gives
\begin{equation}
P_{free,st}^{ud}(v_f)=\frac{1}{\left(D/(\alpha \gamma^{1-\alpha})\right)^{1/\alpha}}L_{\alpha,0}\left(\frac{v_f}{\left(D/(\alpha \gamma^{1-\alpha})\right)^{1/\alpha}}\right).
\end{equation}
This reduces to the Maxwell Boltzmann velocity distribution for $\alpha=2$.
$P_{free}^{ud}(x_{f},T;x_{0},v_0,0)$ and $P_{free}^{ud}(v_f,T;v_0,0)$
for the special case of Cauchy noise (i.e. $\alpha=1$) have been obtained earlier \cite{Seshadri,Poland}.
For $\alpha=2$, we reproduce the expressions given in Chandrasekhar's paper \cite{Chandrasekhar}.

\subsection{Linear Potential}
\label{Free_Levy_under}

The propagator for the Lévy flight under a linear potential of the form $V(x)=-F_0 x$ in the underdamped regime is
\begin{equation}
P_{lin}^{ud}(x_{f},v_f,T;x_{0},v_0,0)=\int_{x(0)=x_{0}}^{x(T)=x_{f}}Dx\int Dp\mbox{ }e^{-D\int_{0}^{T}dt\mbox{ }|p(t)|^{\alpha}}e^{-i\int_{0}^{T}dt\mbox{ }p(t)\left(\frac{\ddot{x}(t)}{\gamma}+\dot{x}(t)-\frac{F_0}{\gamma m}\right)}.\label{free_underdamp}
\end{equation}
Since, $V'(x)$ is independent of the position in this case, the part of the above expression dependent on position is exactly similar to the case of free L\'{e}vy flight. On following the same procedure as we did for free L\'{e}vy flight, we obtain the normalized propagator to be
\begin{eqnarray}
P_{lin}^{ud}(x_{f},v_f,T;x_{0},v_0,0)&&=\frac{1}{4\pi^2\gamma\left(1-e^{-\gamma T}\right)}\int_{-\infty}^{\infty}dp_{0}\int_{-\infty}^{\infty}dp_{T}\mbox{ }e^{-D\int_{0}^{T}dt\mbox{ }\left|\frac{p_0\left(1-e^{-\gamma t}\right)+p_T\left(e^{-\gamma t}-e^{-\gamma T}\right)}{1-e^{-\gamma T}}\right|^{\alpha}}\times\nonumber\\
&&\exp\left[-i\left\{p_T\frac{v_f-\frac{F_0}{m \gamma}}{\gamma}-p_0\frac{v_0-\frac{F_0}{m \gamma}}{\gamma}+\frac{p_0-p_T e^{-\gamma T}}{1-e^{-\gamma T}}\left(x_{f}-x_{0}-\frac{F_0 T}{m \gamma}\right)\right\}\right]\label{lin_underdamp2_1}
\end{eqnarray}
From the above expression it is easy to see that the case of L\'{e}vy flight under a linear potential is equivalent to free L\'{e}vy flight with $v_f\rightarrow\left(v_f-\frac{F_0}{m\gamma}\right)$, $v_0\rightarrow\left(v_0-\frac{F_0}{m\gamma}\right)$ and $x_f\rightarrow \left(x_f-\frac{F_0 T}{m\gamma}\right)$. On making these substitutions in the propagators for free L\'{e}vy flight we can obtain the corresponding propagators for L\'{e}vy flight in a linear potential.

\subsection{Harmonic Potential}

\label{Harmonic_Levy_under}

In the underdamped limit of friction, we analyze Lévy flight in a
harmonic potential of the form $V(x)=\frac{1}{2}\lambda x^{2}$, with
a time-independent force constant. The propagator, $P_{ud}^{har}(x_{f},v_f,T;x_{0},v_0)$,
is
\begin{equation}
P_{ud}^{har}(x_{f},v_f,T;x_{0},v_0)=\int Dx\int Dp\mbox{ }e^{-D\int_{0}^{T}dt\mbox{ }|p(t)|^{\alpha}}e^{-i\int_{0}^{T}dt\mbox{ }p(t)\left(\frac{\ddot{x}(t)}{\gamma}+\dot{x}(t)+\frac{\lambda x(t)}{m\gamma}\right)}.\label{har_underdamp}
\end{equation}
Integrating $\int_{0}^{T}dt\mbox{ }p(t)\{\frac{\ddot{x}(t)}{\gamma}+\dot{x}(t)+\frac{\lambda x(t)}{m\gamma}\}$
in the exponent by parts gives
\begin{eqnarray}
\int_{0}^{T}dt\mbox{ }p(t)\left(\frac{\ddot{x}(t)}{\gamma}+\dot{x}(t)+\frac{\lambda x(t)}{m\gamma}\right) & = & \frac{p_{T}v_f-p_{0}v_0-\dot{p_{T}}x_{f}+\dot{p_{0}}x_{0}}{\gamma}+p_{T}x_{f}-p_{0}x_{0}+\nonumber \\
 &  & \int_{0}^{T}dt\mbox{ }\left(\frac{\ddot{p}(t)}{\gamma}-\dot{p}(t)+\frac{\lambda}{m\gamma}p(t)\right)x(t).
\end{eqnarray}
Using this result in the propagator we obtain,
\begin{eqnarray}
P_{har}^{ud}(x_{f},v_f,T;x_{0},v_0) & = & \int Dp\mbox{ }e^{-D\int_{0}^{T}dt\mbox{ }|p(t)|^{\alpha}}e^{-i\left(\frac{p_{T}v_f-p_{0}v_0-\dot{p_{T}}x_{f}+\dot{p_{0}}x_{0}}{\gamma}+p_{T}x_{f}-p_{0}x_{0}\right)}\times\nonumber \\
 &  & \int Dx\mbox{}e^{-i\int_{0}^{T}dt\mbox{ }\left(\frac{\ddot{p}(t)}{\gamma}-\dot{p}(t)+\frac{\lambda}{m\gamma}p(t)\right)x(t)}\label{har_underdamp1}\\
 & = & \int Dp\mbox{ }e^{-D\int_{0}^{T}dt\mbox{ }|p(t)|^{\alpha}}e^{-i\left(\frac{p_{T}v_f-p_{0}v_0-\dot{p_{T}}x_{f}+\dot{p_{0}}x_{0}}{\gamma}+p_{T}x_{f}-p_{0}x_{0}\right)}\nonumber \\
 &  & \delta\left[\ddot{p}(t)-\gamma\dot{p}(t)+\frac{\lambda}{m}p(t)\right].
\end{eqnarray}
Delta functional implies,
\begin{eqnarray}
 &  & \ddot{p}(t)-\gamma\dot{p}(t)+\frac{\lambda}{m}p(t)=0\label{diff_eq}\\
 & \therefore & p(t)=e^{\frac{\gamma t}{2}}\left(c_{1}\cosh\left(\frac{\beta}{2}t\right)+c_{2}\sinh\left(\frac{\beta}{2}t\right)\right).
\end{eqnarray}
where $\beta=\sqrt{\gamma^{2}-4\frac{\lambda}{m}}$, $c_{1}=p_{0}$
and $c_{2}=\frac{p_{T}e^{-\gamma T}-p_{0}\cosh\left(\frac{\beta}{2}T\right)}{\sinh\left(\frac{\beta}{2}T\right)}$. Though we have shown the working only for the case where $\gamma^2>\frac{4\lambda}{m}$, our analysis is equally valid in general.
Integral over the delta functional will result in the in a double
integral over $p_{T}$ and $p_{0}$. The resultant unnormalized propagator is
\begin{eqnarray}\
\label{most_general_har_1}
 &  & P_{har}^{ud}(x_{f},v_f,T;x_{0},v_0)=\int_{-\infty}^{\infty}dp_{0}\int_{-\infty}^{\infty}dp_{T}\mbox{ }e^{-D\int_{0}^{T}dt\mbox{ }\left|e^{\frac{\gamma t}{2}}\frac{p_{T}e^{-\frac{\gamma T}{2}}\sinh\left(\frac{\beta t}{2}\right)+p_0\sinh\left(\frac{\beta (T-t)}{2}\right)}{\sinh\left(\frac{\beta T}{2}\right)}\right|^{\alpha}}\times\nonumber \\
 &  & e^{-i\left\{p_T\frac{v_f}{\gamma}-p_0\frac{v_0}{\gamma}+\frac{p_T\left(\gamma-\beta\coth\left(\frac{\beta T}{2}\right)\right)+p_0e^{\frac{\gamma T}{2}}\beta\tiny{\mbox{ csch}}\left(\frac{\beta T}{2}\right)}{2\gamma}x_f-\frac{p_0\left(\gamma+\beta\coth\left(\frac{\beta T}{2}\right)\right)-p_Te^{-\frac{\gamma T}{2}}\beta\tiny{\mbox{ csch}}\left(\frac{\beta T}{2}\right)}{2\gamma}x_0\right\}}.
\end{eqnarray}
Upon determining the normalization constant, the propagator is
\begin{eqnarray}
\label{most_general_har_2}
 &  & P_{har}^{ud}(x_{f},v_f,T;x_{0},v_0)=\frac{e^{\frac{\gamma T}{2}}\beta}{8\pi^2 \gamma^2\sinh\left(\frac{\beta T}{2}\right)}\int_{-\infty}^{\infty}dp_{0}\int_{-\infty}^{\infty}dp_{T}\mbox{ }e^{-D\int_{0}^{T}dt\mbox{ }\left|e^{\frac{\gamma t}{2}}\frac{p_{T}e^{-\frac{\gamma T}{2}}\sinh\left(\frac{\beta t}{2}\right)+p_0\sinh\left(\frac{\beta (T-t)}{2}\right)}{\sinh\left(\frac{\beta T}{2}\right)}\right|^{\alpha}}\times\nonumber \\
 &  & e^{-i\left\{\frac{p_T}{\gamma}\left(v_f+\frac{\gamma-\beta\coth\left(\frac{\beta T}{2}\right)}{2}x_f+e^{-\frac{\gamma T}{2}}\beta\tiny{\mbox{ csch}}\left(\frac{\beta T}{2}\right)\frac{x_0}{2}\right)+\frac{p_0}{\gamma}\left(-v_0-\frac{\gamma+\beta\coth\left(\frac{\beta T}{2}\right)}{2}x_0+e^{\frac{\gamma T}{2}}\beta\tiny{\mbox{ csch}}\left(\frac{\beta T}{2}\right)\frac{x_f}{2}\right)\right\}}.
\end{eqnarray}
In order to obtain a more elegant expression for the propagator, we perform a simple change of variables where $\tilde{p}_T=\frac{p_T}{\gamma}$ and $\tilde{p}_0=\frac{p_0 e^{\frac{\gamma T}{2}}}{\gamma\sinh\left(\frac{\beta T}{2}\right)}$ and obtain
\begin{eqnarray}
\label{most_general_har_3}
 &  & P_{har}^{ud}(x_{f},v_f,T;x_{0},v_0)=\frac{\beta}{8\pi^2}\int_{-\infty}^{\infty}d\tilde{p}_0\int_{-\infty}^{\infty}d\tilde{p}_T\mbox{ }e^{-D\int_{0}^{T}dt\mbox{ }\left|\gamma e^{-\frac{\gamma t}{2}}\left(\tilde{p}_0\sinh\left(\frac{\beta t}{2}\right)+\tilde{p}_T\frac{\sinh\left(\frac{\beta (T-t)}{2}\right)}{\sinh\left(\frac{\beta T}{2}\right)}\right)\right|^{\alpha}}\times\nonumber \\
 &  & e^{-i\left\{\tilde{p}_T\left(v_f+\frac{\gamma-\beta\coth\left(\frac{\beta T}{2}\right)}{2}x_f+e^{-\frac{\gamma T}{2}}\beta\tiny{\mbox{ csch}}\left(\frac{\beta T}{2}\right)\frac{x_0}{2}\right)+\tilde{p}_0\left(-v_0e^{-\frac{\gamma T}{2}}\sinh\left(\frac{\beta T}{2}\right)+\frac{\beta x_f}{2}-\frac{\gamma\sinh\left(\frac{\beta T}{2}\right)+\beta\cosh\left(\frac{\beta T}{2}\right)}{2}e^{-\frac{\gamma T}{2}}x_0\right)\right\}}.
 \end{eqnarray}
It is possible to get the stationary distribution by allowing $T\rightarrow \infty$.  Noting that we are considering the case  $\gamma > \beta$, and letting $T\rightarrow \infty$ leads to
\begin{eqnarray}
\label{most_general_har_4}
P_{har,st}^{ud}(x_{f},v_f)&=&\frac{\beta}{8\pi^2}\int_{-\infty}^{\infty}d\tilde{p}_0\int_{-\infty}^{\infty}d\tilde{p}_T\mbox{ }e^{-D\int_{0}^{\infty}dt\mbox{ }\left|\gamma e^{-\frac{\gamma t}{2}}\left(\tilde{p}_0\sinh\left(\frac{\beta t}{2}\right)+\tilde{p}_T e^{-\beta t/2}\right)\right|^{\alpha}}\times\nonumber \\
&& e^{-i\left\{\tilde{p}_T\left(v_f+\frac{\gamma-\beta}{2}x_f\right)+\tilde{p}_0\frac{\beta x_f}{2}\right\}}.
\end{eqnarray}
The above equation implies that $P_{har,st}^{ud}(x_{f},v_f)$ is the two-dimensional Fourier transform of a well behaved function.  This guarantees that the steady state distribution is well-behaved for any $ 2\geqslant \alpha > 0$.

In Eq. (\ref{most_general_har_3}), after performing the coordinate transformation from $\{\tilde{p}_{0},\tilde{p}_{T}\}$ to $\{q_{1},q_{2}\}$ as we did for the free particle, we get
\begin{equation}
\label{most_general_har_4}
P_{har}^{ud}(x_{f},v_f,T;x_{0},v_0)=\frac{\beta}{4\pi^2\alpha}\int_{-\infty}^{\infty}dq_2\int_{0}^{\infty}dq_1\mbox{ }q_1e^{-\Omega_1|q_1|^{\alpha}}\cos\left(\chi_1q_1\right).
\end{equation}
where,
\begin{eqnarray}
\Omega_1&=&D\int_{0}^{T}dt\mbox{ }\left|\gamma e^{-\gamma t/2}\left(\frac{\sinh\left(\frac{\beta (T-t)}{2}\right)}{\sinh\left(\frac{\beta T}{2}\right)}+q_{2}\sinh\left(\frac{\beta t}{2}\right)\right)\right|^{\alpha},\nonumber\\
\chi_1&=&\left(v_f+\frac{\gamma-\beta\coth\left(\frac{\beta T}{2}\right)}{2}x_f+e^{-\frac{\gamma T}{2}}\beta\mbox{ csch}\left(\frac{\beta T}{2}\right)\frac{x_0}{2}\right)\nonumber\\
&&+q_2\left(-v_0e^{-\frac{\gamma T}{2}}\sinh\left(\frac{\beta T}{2}\right)+\frac{\beta x_f}{2}-\frac{\gamma\sinh\left(\frac{\beta T}{2}\right)+\beta\cosh\left(\frac{\beta T}{2}\right)}{2}e^{-\frac{\gamma T}{2}}x_0\right).
\end{eqnarray}
For $\alpha=1$, the integral over $q_1$ can be performed exactly resulting in the expression
\begin{equation}
\label{most_general_har_5}
P_{har}^{ud}(x_{f},v_f,T;x_{0},v_0)=\frac{\beta}{4\pi^2}\int_{-\infty}^{\infty}dq_2\mbox{ }\frac{\Omega_1^2-\chi_1^2}{\left(\Omega_1^2+\chi_1^2\right)^2}.
\end{equation}
This enables us to calculate the time evolution of the probability distribution easily (see Fig. 1).  For other values of $\alpha$, for which the integral over $q_1$ cannot be performed exactly, we expand the cosine function in Eq. (\ref{most_general_har_4}) as a sum and perform the integral over $q_1$. This results in
\begin{equation}
\label{most_general_har_sum}
P_{har}^{ud}(x_{f},v_f,T;x_{0},v_0)=\frac{\beta}{4\pi^2\alpha}\int_{-\infty}^{\infty}dq_2\sum_{n=0}^{\infty}\frac{(-1)^n}{(2n)!}
\mbox{ }\Gamma\left(\frac{2(n+1)}{\alpha}\right)\frac{\chi_1^{2n}}{\Omega_1^{2(n+1)/\alpha}}.
\end{equation}
This sum does not converge for $\alpha<1$.

Specific forms of the propagator $P_{har}^{ud}(x_{f},T;x_{0},v_0,0)$
and $P_{har}^{ud}(v_f,T;x_{0},v_0,0)$ can be obtained from
$P_{har}^{ud}(x_{f},v_f,T;x_{0},v_0,0)$ by integrating
over $v_f$ and $x_{f}$ respectively. $P_{har}^{ud}(x_{f},T;x_{0},v_0,0)$
is
\begin{equation}
\label{position_har_ud}
P_{har}^{ud}(x_{f},T;x_{0},v_0,0)=\int_{-\infty}^{\infty}dv_f\mbox{ }P_{har}^{ud}(x_{f},v_f,T;x_{0},v_0,0)=\frac{1}{\Omega_2^{1/\alpha}}L_{\alpha,0}\left(\frac{\chi_2}{\Omega_2^{1/\alpha}}\right)
\end{equation}
where,
\begin{eqnarray}
\Omega_2&=&D\int_{0}^{T}dt\mbox{ }\left(\frac{2\gamma}{\beta}e^{-\gamma t/2}\sinh\left(\frac{\beta}{2}t\right)\right)^{\alpha}=D\frac{\gamma^{\alpha}}{\beta^{\alpha+1}}B\left(1-e^{-\beta T};\alpha+1,\frac{\alpha}{2}\left(\frac{\gamma}{\beta}-1\right)\right)\nonumber\\
\chi_2&=&x_{f}-\left(\cosh\left(\frac{\beta T}{2}\right)+\frac{\gamma}{\beta}\sinh\left(\frac{\beta T}{2}\right)\right)e^{\frac{-\gamma T}{2}}x_{0}-\frac{2v_0}{\beta}e^{\frac{-\gamma T}{2}}\sinh\left(\frac{\beta T}{2}\right).
\end{eqnarray}
The stationary distribution for the propagator $P_{har}^{ud}(x_{f},T;x_{0},v_0,0)$ is
\begin{equation}
P_{har,st}^{ud}(x_{f})=\int_{-\infty}^{\infty}dv_f\mbox{ }P_{har}^{ud}(x_{f},v_f,T;x_{0},v_0,0)=\frac{1}{\Omega_{2,st}^{1/\alpha}}L_{\alpha,0}\left(\frac{\chi_{2,st}}{\Omega_{2,st}^{1/\alpha}}\right)
\end{equation}
where,
\begin{eqnarray}
\Omega_{2,st}&=&D\int_{0}^{\infty}dt\mbox{ }\left(\frac{2\gamma}{\beta}e^{-\gamma t/2}\sinh\left(\frac{\beta}{2}t\right)\right)^{\alpha}=D\frac{\gamma^{\alpha}}{\beta^{\alpha+1}}B\left(\frac{\alpha}{2}\left(\frac{\gamma}{\beta}-1\right),\alpha+1\right),\nonumber\\
\chi_{2,st}&=&x_{f}.
\end{eqnarray}
Also,
\begin{equation}
\label{velocity_har_ud}
P_{har}^{ud}(v_f,T;x_{0},v_0,0)=\int_{-\infty}^{\infty}dx_{f}\mbox{ }P_{har}^{ud}(x_{f},v_f,T;x_{0},v_0,0)=\frac{\beta}{2\Omega_3^{1/\alpha}}L_{\alpha,0}\left(\frac{\chi_3}{\Omega_3^{1/\alpha}}\right)
\end{equation}
where,
\begin{eqnarray}
\Omega_3&=&D\int_{0}^{T}dt\mbox{ }\left|\gamma e^{-\gamma t/2}\left(\cosh\left(\frac{\beta}{2}t\right)-\frac{\gamma}{\beta}\sinh\left(\frac{\beta}{2}t\right)\right)\right|^{\alpha},\nonumber\\
\chi_3&=&v_f+\left(\frac{\gamma}{\beta}\sinh\left(\frac{\beta T}{2}\right)-\cosh\left(\frac{\beta T}{2}\right)\right)e^{-\gamma T/2}v_0+
\left(\frac{(\gamma^2-\beta^2)\sinh\left(\frac{\beta T}{2}\right)}{2\beta}\right)e^{-\gamma T/2} x_0.
\end{eqnarray}
The stationary distribution for the propagator $P_{har}^{ud}(x_{f},T;x_{0},v_0,0)$ is
\begin{equation}
P_{har,st}^{ud}(v_f)=\int_{-\infty}^{\infty}dx_{f}\mbox{ }P_{har}^{ud}(x_{f},v_f,T;x_{0},v_0,0)=\frac{\beta}{2\Omega_{3,st}^{1/\alpha}}L_{\alpha,0}\left(\frac{\chi_{3,st}}{\Omega_{3,st}^{1/\alpha}}\right)
\end{equation}
where,
\begin{eqnarray}
\Omega_{3,st}&=&D\int_{0}^{\infty}dt\mbox{ }\left|\gamma e^{-\gamma t/2}\left(\cosh\left(\frac{\beta}{2}t\right)-\frac{\gamma}{\beta}\sinh\left(\frac{\beta}{2}t\right)\right)\right|^{\alpha},\nonumber\\
\chi_{3,st}&=&v_f.
\end{eqnarray}
For $\alpha=2$, we obtain the results of Chandrasekhar \cite{Chandrasekhar}.

The propagators in the case where $\gamma^2<\frac{4\lambda}{m}$ can be obtained from the expressions derived for $\gamma^2>\frac{4\lambda}{m}$ by replacing $\cosh[\frac{\beta T}{2}]$ with $\cos[\frac{\tilde{\beta} T}{2}]$ and $\frac{1}{\beta}\sinh[\frac{\beta T}{2}]$ with $\frac{1}{\tilde{\beta}}\sin[\frac{\tilde{\beta} T}{2}]$ where, $\tilde{\beta}=\sqrt{\frac{4\lambda}{m}-\gamma^2}$. For $\gamma^2=\frac{4\lambda}{m}$ case, the appropriate propagators can be obtained by replacing $\cosh[\frac{\beta T}{2}]$ with $1$ and $\frac{1}{\beta}\sinh[\frac{\beta T}{2}]$ with $\frac{T}{2}$ in the expressions obtained for $\gamma^2>\frac{4\lambda}{m}$. The expectation value of the energy of a particle executing L\'{e}vy flight in a harmonic potential in the underdamped limit will be given by $\frac{1}{2}m \langle v_f^2\rangle+\frac{1}{2}\lambda \langle x_f^2\rangle$. Since both the position and the velocity distributions are L\'{e}vy stable, as given in Eq. (\ref{position_har_ud}) and (\ref{velocity_har_ud}) respectively, both $\langle x_f^2\rangle$ and $\langle v_f^2\rangle$ diverge.

Phase space distributions for L\'{e}vy flights under a harmonic potential have been explored previously (see references \cite{sokolov1,cao}). In the case usual Brownian motion, $x_f$ and $v_f$ are uncorrelated and one obtains elliptical contours for the phase space distribution. However, for L\'{e}vy flights this is no longer true - an inhomogeneous stationary phase space distribution was observed \cite{cao,sokolov1}. Since, our method gives exact expressions for propagators at any time, we can study the time evolution of the phase space distribution, while the previous papers were concerned only with the steady state distribution function. In Fig. (1(a),1(b),1(c),1(d)), we show the time evolution of the phase space distribution for $\alpha=1$ with the initial conditions, $v_f=0$ and $x_f=0$. In Fig. (1(e)), we also show the stationary distribution calculated simply as the  product of $P_{har}^{ud}(x_f,T;v_0,x_0,0)$ and $P_{har}^{ud}(v_f,T;v_0,x_0,0)$ in the large $T$ limit. We observe that there is correlation between $x_f$ and $v_f$ which does not die off as $T\rightarrow \infty$.  Our findings are consistent with the inhomogeneity in phase space distribution reported in the literature \cite{sokolov1,cao}.

\begin{figure}[h]
\label{figure1}
\centering
\subfigure[\mbox{ }$P^{ud}_{har}(v_f,x_f,T;0,0,0)$ at T=1]{
\includegraphics [scale=0.45] {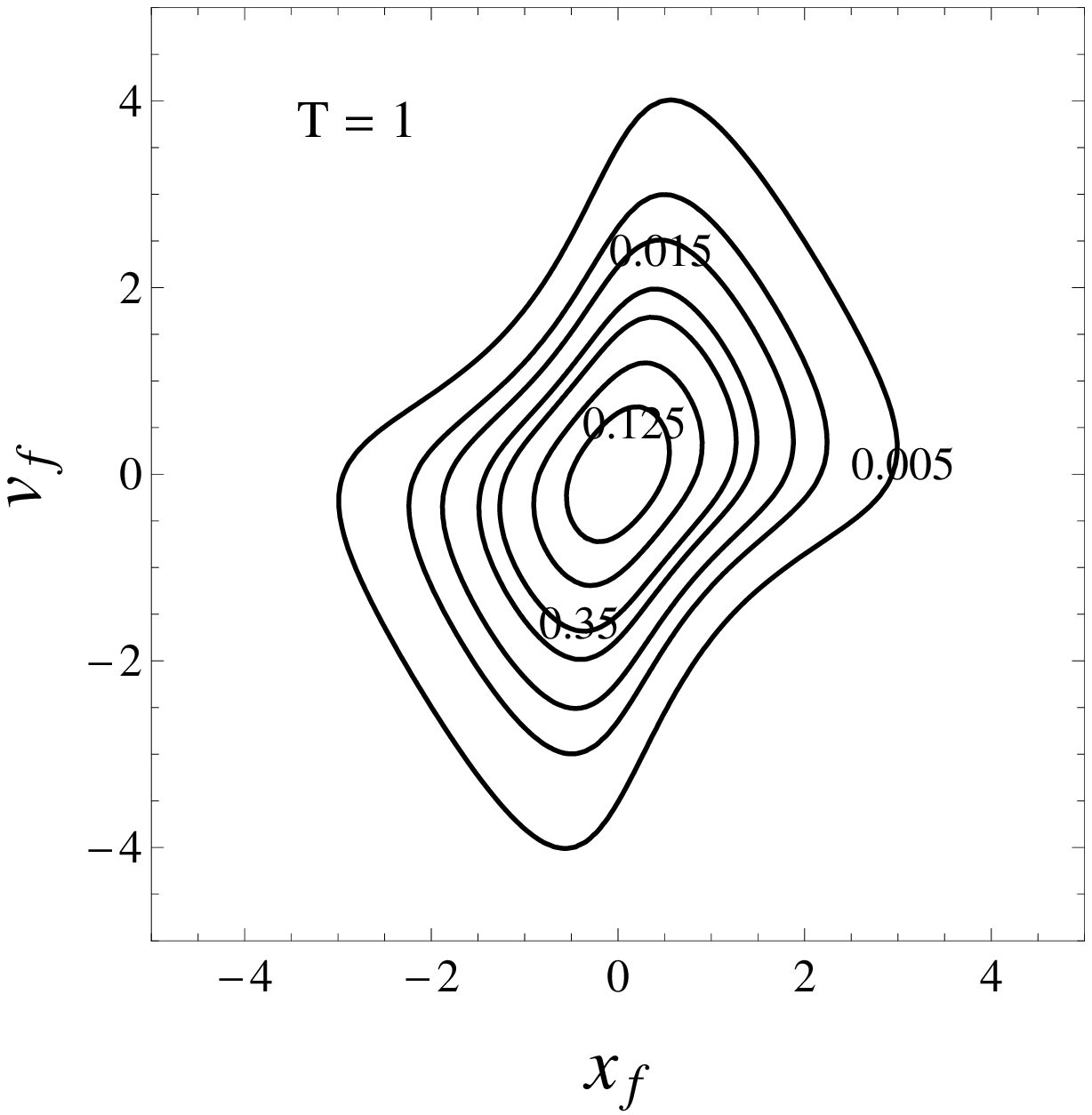}}
\subfigure[\mbox{ }$P^{ud}_{har}(v_f,x_f,T;0,0,0)$ at T=3]{
\includegraphics [scale=0.45] {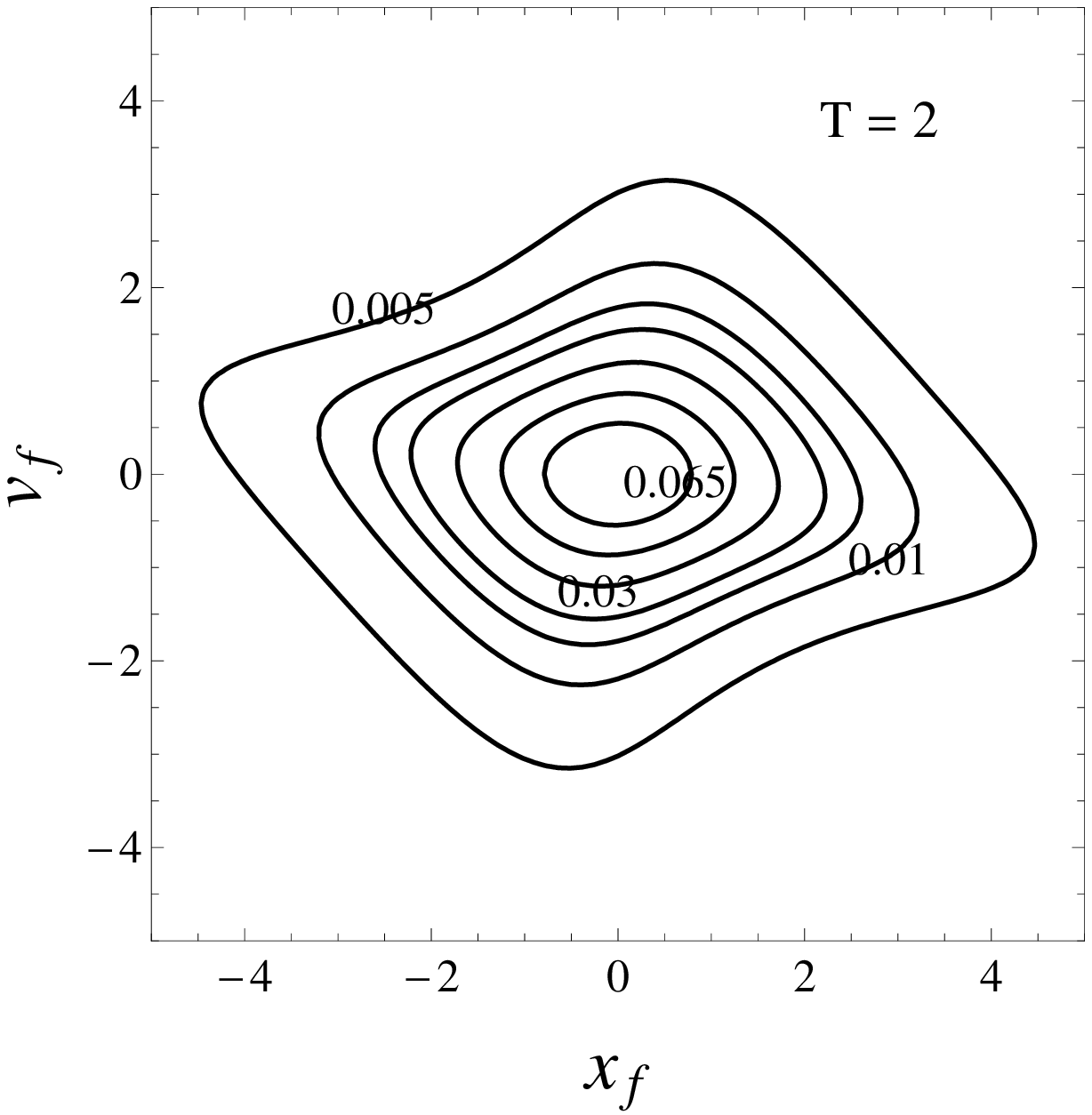}}
\subfigure[\mbox{ }$P^{ud}_{har}(v_f,x_f,T;0,0,0)$ at T=5]{
\includegraphics [scale=0.45] {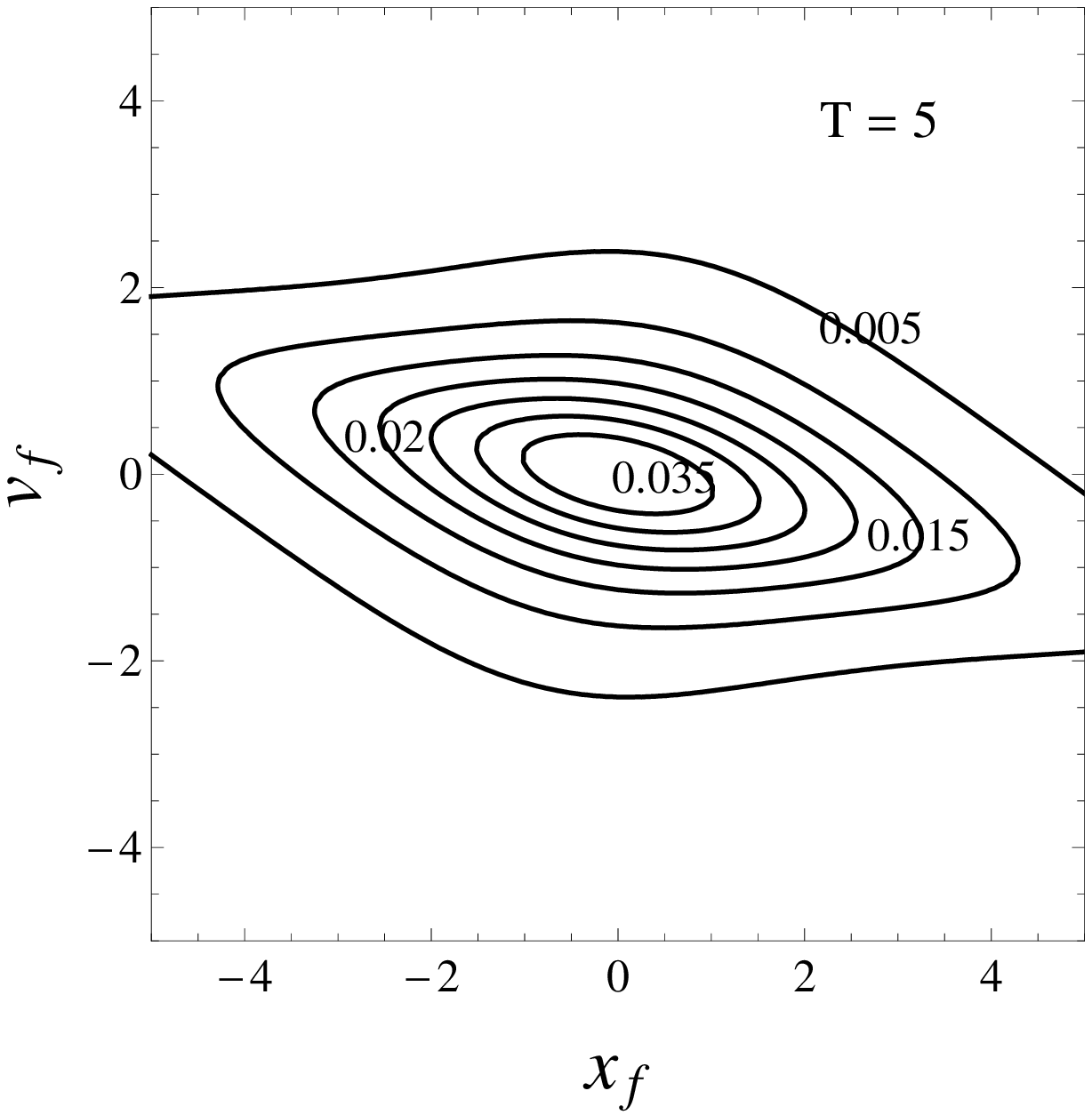}}
\subfigure[\mbox{ }$P^{ud}_{har}(v_f,x_f,T;0,0,0)$ at long time]{
\includegraphics [scale=0.45] {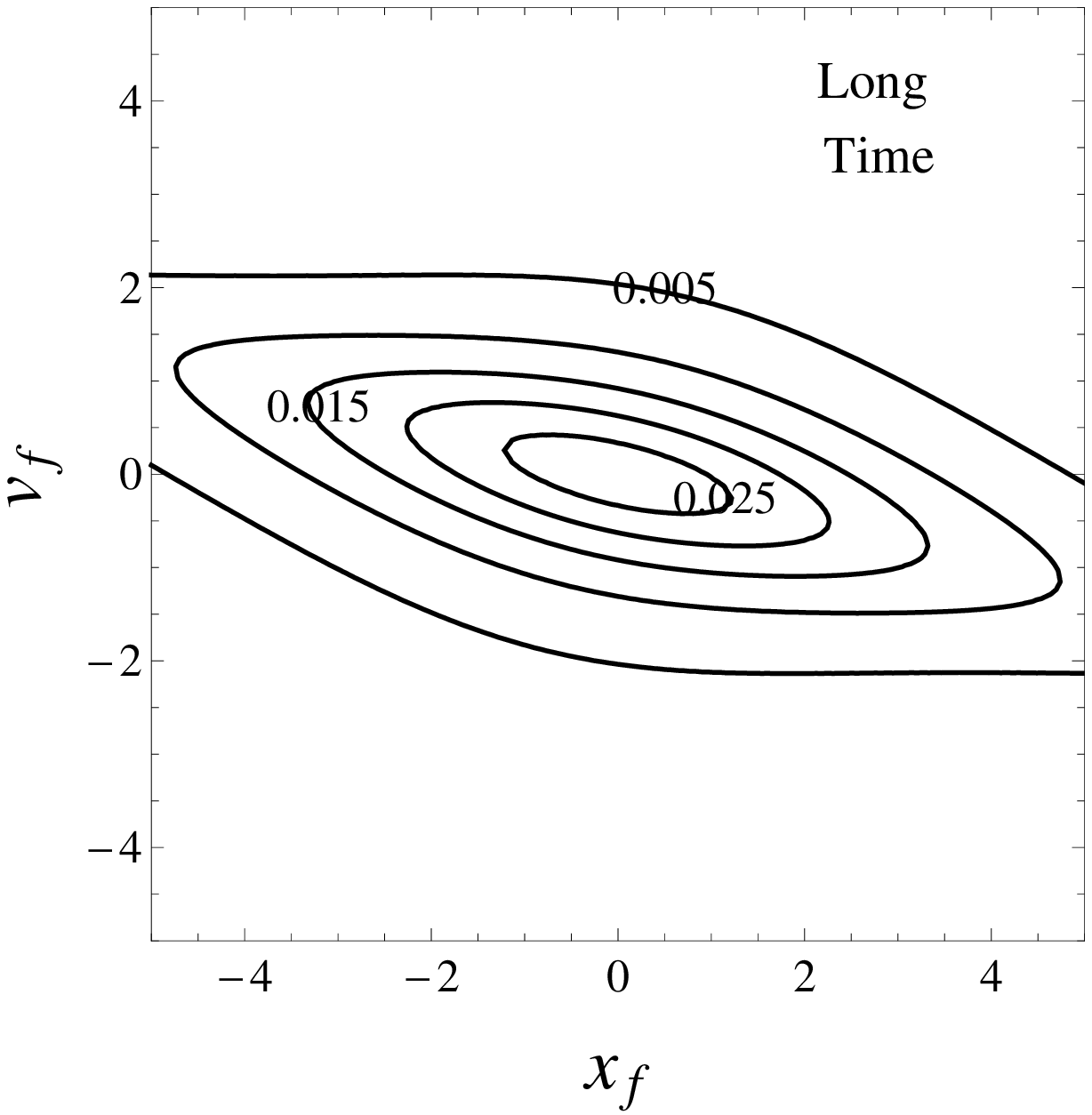}}
\subfigure[\mbox{ }Product of the Stationary distributions $P^{ud}_{har}(x_f)$ and $P^{ud}_{har}(v_f)$]{
\includegraphics [scale=0.45] {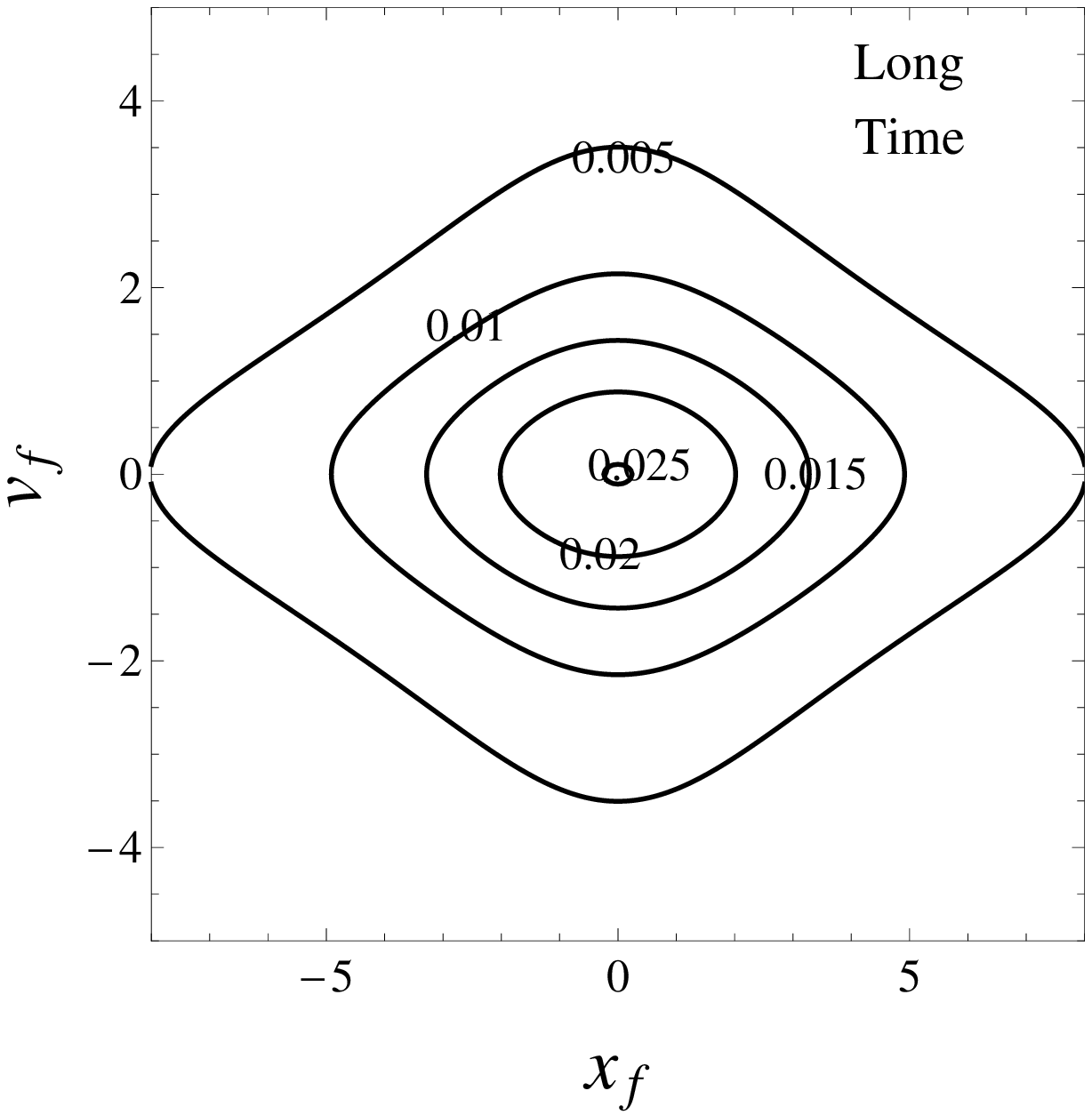}}
\caption{The figures (a)-(d) show the time evolution of $P^{ud}_{har}(v_f,x_f,T;v_0,x_0,0)$ starting with $v_0=0$ and $x_0=0$. Figure (e) shows the long time behavior of the product of $P^{ud}_{har}(x_f,T;v_0,x_0,0)$ and $P^{ud}_{har}(v_f,T;v_0,x_0,0)$. (a) $P^{ud}_{har}(v_f,x_f,T;0,0,0)$ at T=1; (b) $P^{ud}_{har}(v_f,x_f,T;0,0,0)$ at T=3; (c) $P^{ud}_{har}(v_f,x_f,T;0,0,0)$ at T=5; (d) (a) $P^{ud}_{har}(v_f,x_f,T;0,0,0)$ at long time; (e) Product of the Stationary distributions $P^{ud}_{har}(x_f)$ and $P^{ud}_{har}(v_f)$.}
\end{figure}

\section{Conclusions}

We have shown that Hamiltonian path integrals offer an appealing way
to study Lévy flights in both the overdamped and underdamped limits
of friction. Though some results are already
available, the method adopted to obtain them is interesting.
In addition, they enable us to obtain expressions for several propagators for Lévy flights, which seem difficult to obtain using other methods.
\section{Acknowledgements}
We thank Professor R. Friedrich for sending us a copy of reference \cite{Fredriech} and an anonymous referee for pointing out reference \cite{BarkaiJStatPhys2010}.   The work of K.L. Sebastian was supported by the Department of Science and Technology, Govt. of India by the J.C. Bose fellowship program. Deepika Janakiraman acknowledges CSIR, India for support though a scholarship.
\section{Appendix}

\subsection{Fractional Fokker-Planck Equation from the path integral}

Here we derive the FFPE starting from the discretized of the path integral as given in Eq. (\ref{eq:DiscretizedVersion}). Let $P(x,t+\Delta t;x_0,0)$ be the probability of finding the particle at $x$ at the time $t+\Delta t$ given that it started at $x_0$ at the time $t=0$.  Then the path integral prescription of Eq. (\ref{eq:DiscretizedVersion}) enables us to calculate this  probability density in terms of $P(x',t;x_0,0)$, the probability distribution at the time $t$, as
\begin{equation}
P(x,t+\Delta t;x_0,0)=\frac{1}{2\pi}\int^{\infty}_{-\infty}dp\int^{\infty}_{-\infty}dx'\mbox{ }e^{-D |p|^{\alpha}\Delta t} e^{ip\left\{(x-x')+\frac{V'(x')}{\gamma m}\Delta t\right\}}P(x',t;x_0,0).
\end{equation}
Since $\Delta t$ is infinitesimally small, we can expand the exponential $e^{\left(-D |p|^{\alpha} +i p \frac{V'(x')}{\gamma m}\right)\Delta t}$  up to linear order in $\Delta t$ and obtain,
\begin{equation}
\label{prove_FFPE_2}
P(x,t+\Delta t;x_0,0)=\frac{1}{2\pi}\int^{\infty}_{-\infty}dp\int^{\infty}_{-\infty}dx'\mbox{ }\left(1-D |p|^{\alpha}\Delta t +i p \Delta t \frac{V'(x')}{\gamma m}\right) e^{ip(x-x')}P(x',t;x_0,0).
\end{equation}
Defining the operator $\left (-\frac{\partial^2}{\partial x^2}\right)^{\alpha/2} $ by
\begin{equation}
D\left(-\frac{\partial^2}{\partial x^2}\right)^{\alpha/2}P(x,t;x_0,0) =\frac{1}{2\pi}\int^{\infty}_{-\infty}dp\int^{\infty}_{-\infty}dx' D |p|^{\alpha} e^{ip(x-x')}P(x',t;x_0,0),
\end{equation}
 we can write Eq. (\ref{prove_FFPE_2}) as
\begin{equation}
\label{prove_FFPE_2_RHS}
P(x,t+\Delta t;x_0,0)=P(x,t;x_0,0)+\Delta t\left(-D\left(-\frac{\partial^2}{\partial x^2}\right)^{\alpha/2}+\frac{\partial}{\partial x}\frac{V'(x)}{\gamma m}\right)P(x,t;x_0,0).
\end{equation}
From the above equation, we obtain the FFPE given in Eq. (\ref{Pot FFPE}) which is
\begin{equation}
\frac{\partial P(x,t;x_0,0)}{\partial t}=\left(-D\left(-\frac{\partial^2}{\partial x^2}\right)^{\alpha/2}+\frac{\partial}{\partial x} \frac{V'(x)}{\gamma m}\right)P(x,t;x_0,0).
\end{equation}

\subsection{Transformation of coordinates from $\{p_{0},p_{T}\}$ to $\{q_{1},q_{2}\}$}

The propagators $P_{free}^{ud}(x_{f},v_f,T;x_{0},v_0,0)$
and $P_{har}^{ud}(x_{f},v_f,T;x_{0},v_0,0)$ can be
written in a generalized form as
\begin{equation}
\label{Pgen}
P_{gen}=\int_{-\infty}^{\infty}dp_{T}\int_{-\infty}^{\infty}dp_{0}\mbox{ }e^{-\int_{0}^{T}dt\mbox{  }|a(t)p_{T}+b(t)p_{0}|^{\alpha}}e^{-i(gp_{T}+hp_{0})}
\end{equation}
which shows the formal dependence of $p_{0}$ and $p_{T}$. We will
not work out the steps explicitly for $P_{free}^{ud}(x_{f},v_f,T;x_{0},v_0,0)$
and $P_{har}^{ud}(x_{f},v_f,T;x_{0},v_0,0)$ but only
for $P_{gen}$. The propagators of interest can be obtained from $P_{gen}$.
Using the transformation $\{p_{0},p_{T}\}$ to $\{q_{1},q_{2}\}$
with $p_{T}=q_{1}$ and $p_{0}=q_{1}q_{2}$, we get
\begin{equation}
\label{Pgen1}
P_{gen}=2\int_{-\infty}^{\infty}dq_{2}\int_{0}^{\infty}dq_{1}\mbox{ }q_{1}e^{-|q_{1}|^{\alpha}\int_{0}^{T}dt\mbox{ }|a(t)+b(t)q_{2}|^{\alpha}}\cos((g+hq_{2})q_{1}).
\end{equation}
Upon expanding the cosine series and performing the integral over $q_1$, we get
\begin{equation}
\label{Pgen2}
P_{gen}=\frac{2}{\alpha}\int_{-\infty}^{\infty}dq_{2}\mbox{ }\sum_{n=0}^{\infty}\frac{(-1)^n}{(2n)!}\mbox{ }\Gamma\left(\frac{2(n+1)}{\alpha}\right)\frac{(g+hq_{2})^{2n}}{\left(\int_{0}^{T}dt\mbox{ }
|a(t)+b(t)q_{2}|^{\alpha}\right)^{2(n+1)/\alpha}}.
\end{equation}
Note that the above sum can be evaluated exactly for $\alpha=1$ or $2$ and lead to Cauchy or
exponential functions. Further, the sum does not converge if $\alpha<1$.

\subsection{Calculation of the propagator $P_{free}^{ud}(x_{f},T;x_{0},v_0,0)$}

\label{Appendix_B} Our procedure enables us to calculate the most
general propagator in the underdamped regime for both free Lévy flight
and that under a linear and a harmonic potential. From these, we showed results
for more specific propagators such as $P_{free}^{ud}(x_{f},T;x_{0},v_0,0)$,
$P_{free}^{ud}(v_f,T;v_0,0)$, $P_{har}^{ud}(x_{f},T;x_{0},v_0,0)$,
and $P_{har}^{ud}(v_f,T;x_{0},v_0,0)$.
Here, in the appendix, we will show the steps involved in calculating
$P_{free}^{ud}(x_{f},T;x_{0},v_0,0)$. The rest
of the propagators can be obtained in a similar fashion.
\begin{eqnarray}
P_{free}^{ud}(x_{f},T;x_{0},v_0,0)& = & \int_{-\infty}^{\infty}dv_f\mbox{ }P_{free}^{ud}(x_{f},v_f,T;x_{0},v_0,0)\nonumber \\
&=&\frac{1}{4\pi^2\gamma\left(1-e^{-\gamma T}\right)}\int_{-\infty}^{\infty}dp_{0}\int_{-\infty}^{\infty}dp_{T}\mbox{ }e^{-D\int_{0}^{T}dt\mbox{ }\left|\frac{p_0\left(1-e^{-\gamma t}\right)-p_T\left(e^{-\gamma T}-e^{-\gamma t}\right)}{1-e^{-\gamma T}}\right|^{\alpha}}\times\nonumber\\
&&e^{-i\left\{-p_0\frac{v_0}{\gamma}+\frac{p_0-p_T e^{-\gamma T}}{1-e^{-\gamma T}}(x_{f}-x_{0})\right\}}\int_{-\infty}^{\infty}dv_f\mbox{ }e^{-ip_T\frac{v_f}{\gamma}}\nonumber\\
&=&\frac{1}{2\pi\left(1-e^{-\gamma T}\right)}\int_{-\infty}^{\infty}dp_{0}\int_{-\infty}^{\infty}dp_{T}\mbox{ }e^{-D\int_{0}^{T}dt\mbox{ }\left|\frac{p_0\left(1-e^{-\gamma t}\right)-p_T\left(e^{-\gamma T}-e^{-\gamma t}\right)}{1-e^{-\gamma T}}\right|^{\alpha}}\times\nonumber\\
&&e^{-i\left\{-p_0\frac{v_0}{\gamma}+\frac{p_0-p_T e^{-\gamma T}}{1-e^{-\gamma T}}(x_{f}-x_{0})\right\}}\delta(p_T)
\end{eqnarray}
Performing the integration over the delta function we obtain
\begin{eqnarray}
P_{free}^{ud}(x_{f},T;x_{0},v_0,0)& = &\frac{1}{2\pi\left(1-e^{-\gamma T}\right)}\int_{-\infty}^{\infty}dp_{0}\mbox{ }e^{-D\int_{0}^{T}dt\mbox{ }\left|p_0\frac{1-e^{-\gamma t}}{1-e^{-\gamma T}}\right|^{\alpha}}e^{i p_0\left\{\frac{v_0}{\gamma}-\frac{x_{f}-x_{0}}{1-e^{-\gamma T}}\right\}}\nonumber\\
&=&\frac{1}{\left(D\int_{0}^{T}dt\mbox{ }\left(1-e^{-\gamma t}\right)^{\alpha}\right)^{1/\alpha}}L_{\alpha,0}\left(\frac{\frac{v_0(1-e^{-\gamma T})}{\gamma}-(x_{f}-x_{0})}{\left(D\int_{0}^{T}dt\mbox{ }\left(1-e^{-\gamma t}\right)^{\alpha}\right)^{1/\alpha}}\right).
\end{eqnarray}

\end{document}